\documentclass[aps,prd,superscriptaddress,showpacs,showkeys,floatfix,preprintnumbers,nofootinbib,letterpaper]{revtex4-2}
\usepackage[Export]{adjustbox}
\usepackage{setspace}
\usepackage{verbatim}
\usepackage{tikz}
\usetikzlibrary {arrows.meta}
\usepackage{color}
\usepackage{amssymb}
\usepackage{pgfplots}
\usepackage{amsmath,amssymb}
\usepackage{mathtools}
\usepackage{hyperref}

\usepackage{graphicx}  
\usepackage{dcolumn}   
\usepackage{bm}        
\usepackage{amssymb}   
\usepackage{physics}
\usepackage{mathtools}
\usepackage{ulem}
\usepackage{placeins}

\definecolor{Rocco}{rgb}{0.0, 0.5, 0.0}
\definecolor{Resolved?}{rgb}{0.5, 0.0, 0.5}

\newcommand{\DU}{\bar{D}}
\newcommand{\talpha}{\tilde{\alpha}}

\def\Tr{{\rm Tr}}
\def\half{{\textstyle{1\over 2}}}
\def\A{{\mathcal A}}
\def\D{{\mathcal D}}
\def\G{{\mathcal G}}
\def\P{{\mathcal P}}
\def\vx{\vec x}
\allowdisplaybreaks

\begin{document}
\title{Goldstone theory of gluon saturation in high energy QCD:\\ From color glass to color superfluid condensate}
\author{V. P. Nair}
\affiliation{Physics Department, City College of New York, CUNY, New York, NY, 10031}
\author{Raju Venugopalan}
\affiliation{Physics Department, Brookhaven National Laboratory, Upton, NY 11973, USA}
\affiliation{CFNS, Department of Physics and Astronomy, Stony Brook University, Stony Brook, NY 11794, USA}
\affiliation{Higgs Center for Theoretical Physics, The University of Edinburgh, Edinburgh, EH9 3FD, Scotland, UK}
\begin{abstract}
In the high energy Regge asymptotics of QCD, the rapid overpopulation of gluons leads to gluon saturation, characterized by nonperturbative semi-classical lumpy configurations of size $1/Q_S$, where $Q_S\gg \Lambda_{\rm QCD}$ is a dynamical semi-hard saturation scale. We discuss here a generalization of the Color Glass Condensate (CGC) framework describing this phenomenon that is obtained by  carefully treating the breaking of global symmetries by the condensate. 
In particular, we show that the breaking of a global subgroup of color results in a Goldstone effective field theory (EFT) whose dynamics is characterized by a Goldstone decay constant $f_G = \sqrt{N} Q_S$, where $N=O(1/\alpha_S)$ is the phase space occupancy of the condensate, which is large for $\alpha_S(Q_S)\ll 1$. Further, the coupling of these colored Goldstones to collinear gluon modes leads to a depletion of the condensate, thereby modifying the static shockwave scattering framework of the CGC on timescales controlled by $f_G$. The CGC EFT on these longer time scales is a novel color superfluid. Low frequency modes with $\omega\leq \alpha_S Q_S$ in an external probe of the CGC are sensitive to superfluid dynamics, with interesting physical consequences for deeply inelastic scattering and ultrarelativistic nuclear collisions.

\pacs{11.15.Ha}
\end{abstract}

\date{\today}

\maketitle
\section{Introduction}

The focus of this work is the generalization of 
the effective field theory (EFT) framework of the Color Glass
Condensate (CGC)~\cite{Iancu:2003xm,Gelis:2010nm,Kovchegov:2012mbw,Raj:2025hse}, which describes the phenomenon of gluon saturation in QCD at high energies~\cite{Gribov:1983ivg,Mueller:1985wy,Mueller:2026min}.  In the language of deep inelastic scattering (DIS), gluon saturation occurs in Regge asymptotics where the Bjorken variable $x_{\rm Bj}\sim Q^2/s\rightarrow 0$ for fixed resolution $Q^2 \gg \Lambda_{\rm QCD}^2$ and squared center-of-mass energy $s\rightarrow \infty$.\footnote{More generally, the high energy regime is $s\rightarrow \infty$, $s/|t|\rightarrow \infty$, $|t|\gg \Lambda_{\rm QCD}$, where $s=(p_1+p_2)^2$ and $t=(p_1-p_3)^2$ are the Mandelstam invariants for the elastic scattering process with incoming and outgoing momenta of the particles, respectively, $p_1+p_2\rightarrow p_3+p_4$.} 
For the purposes of this discussion, we will equate, as in Feynman's parton model, $x_{\rm Bj}\sim x$, where $x$ is the parton momentum fraction in a high energy hadron.\footnote{The relation between the two variables is process-dependent but can be worked out systematically in perturbation theory.} As in any EFT, the CGC relies on a Born-Oppenheimer type separation.\footnote{There is an isomorphism between fast and slow lightcone momenta $k^+$ in the Poincar\'{e} group on the lightfront, and that of heavy and light masses, respectively, in the Galilean sub-group of 2-D quantum mechanics~\cite{Susskind:1967rg,Bardakci:1968zqb,Bjorken:1970ah,Venugopalan:1998zd}.} In this case, it is between fast (large $x$) degrees of freedom that are static on the lightfront and slow (small $x$) ``wee parton" dynamical modes~\cite{McLerran:1993ka,McLerran:1993ni}. 

Before we dive into the physics of gluon saturation and the CGC EFT, it is useful to consider how this phenomenon arises in the first place. The underlying problem is a first principles description of $2\rightarrow 2+n$ scattering at collider energies when $n\gg 1$. In Regge asymptotics, one anticipates the problem should be under control because the coupling $\alpha_S(Q^2)\ll 1$ and one can therefore employ perturbation theory. As we will discuss, this is not the case, and nonperturbative methods are necessary even for arbitrary small coupling if $x$ is sufficiently small.

The perturbative framework for $2\rightarrow 2+n$ scattering was developed by BFKL~\cite{Kuraev:1976ge,Kuraev:1977fs,Balitsky:1978ic}; this consists of identifying the largest logarithms $\alpha_S\ln(1/x)$ at each order in perturbation theory, and resumming these contributions to all orders. In this leading logarithmic approximation (LLx) resumming all $(\alpha_S\ln(1/x))^n$ contributions, the corresponding Feynman diagrams can be organized into a ladder-like structure. The $n$ rungs of the ladder correspond to the $n$ emitted gluons in the $2\rightarrow 2+n$ scattering amplitude, the vertices at both ends are so-called effective Lipatov vertices, and the propagators along the sides of the ladder are composite reggeized gluons. The total cross-section at this LLx accuracy is the imaginary part of the color singlet projection of the amplitude, corresponding to the $t$-channel exchange of a pomeron. This object is a key element of reggeon field theory introduced prior to QCD to describe the growth of total cross-sections at high energies~\cite{Forshaw:1997dc,Donnachie:2013xia}. The ladder structure of the BFKL pomeron is captured by the BFKL renormalization group (RG) equation describing the change of the amplitude with increasing rapidity $Y$($=\ln(1/x)$); at LLx, its solution gives a rapidly growing cross-section $\sigma(s)\sim s^{0.5}$ for $\alpha_S=0.2$. 

The BFKL construction has been extended to next-to-leading logarithmic accuracy (NLLx) (resumming $\alpha_S^{n+1} \ln^n(1/x)$ logs)~\cite{Fadin:1998py,Ciafaloni:1998gs}; extracting its predictive power for the cross-section requires in addition a further resummation of large double logarithms in $Q^2$~\cite{Salam:1998tj,Ciafaloni:1999yw,Ciafaloni:2003rd,Forte:2009wh}. These higher order corrections tame the rapid LLx growth of the cross-section but the energy dependence nevertheless remains strong at HERA DIS energies\footnote{For a quantitative comparison of BFKL to HERA DIS data, see for instance~\cite{Ball:2017otu}.}, with $\sigma(s)\sim s^{0.3}$. Towards NNLLx accuracy, there has been impressive recent progress~\cite{DelDuca:2022skz} in computing two-loop contributions to the Lipatov vertex~\cite{Abreu:2024xoh} and three-loop contributions to $t$-channel propagators~\cite{Falcioni:2021dgr,Fadin:2024eyf}. 

Despite its theoretical and phenomenological success, solutions of the BFKL RG equations exhibit the unwelcome property, both at LLx and at NLLx, that gluons with large transverse momentum diffuse into the infrared at sufficiently large rapidities. This results in  increasing sensitivity to nonperturbative physics signaling the breakdown of the operator product expansion (OPE) in this framework~\cite{Mueller:1996hm}. Further, the rapid growth of parton phase space distributions suggests that higher twist (power suppressed) contributions become uncontrollably large at small $x$~\cite{Gribov:1983ivg,Mueller:1985wy}. Both of these effects suggest a breakdown of perturbation theory in Regge asymptotics. 

Gluon saturation neatly resolves both of these problematic conceptual features of BFKL dynamics. Simply put, it is the statement that gluon field strengths reach a maximal phase space occupancy of $O(1/\alpha_S)$ for transverse momentum modes $k_\perp\sim Q_S$, where $Q_S(x)\gg \Lambda_{\rm QCD}$ is an emergent semi-hard ``saturation scale" in the theory.\footnote{A nuanced recent discussion of gluon phase space occupancies for modes $\Lambda_{\rm QCD}\ll k_\perp \ll Q_S$ can be found in \cite{Mueller:2026fee}. These occupancies can be even larger but only resolve limited phase space.}. Equivalently, in the S-matrix picture, gluon saturation corresponds to small localized ``lumpy" gauge field configurations of size $1/Q_S(x)$ in a hadron, which a probe scatters off with unit probability. With increasing energy, the lumps get smaller but there are many more of them, with the result that the hadron gets increasingly opaque or ``black".\footnote{Gluon saturation however is not sufficient to saturate the Froissart bound, which requires nonperturbative physics of the strong coupling variety~\cite{Kovner:2001bh,Kovner:2002yt}. } In Regge asymptotics, $Q_S$ is the largest scale in the problem, suggesting that $\alpha_S\equiv \alpha_S(Q_S)\ll 1$. 

The saturated lumps are nonperturbative semi-classical gauge field configurations because of their large $O(1/\alpha_S)$ phase space occupancy: What is remarkable is that their many-body emergent dynamics can be explored employing weak coupling methods. One consequence, as we will discuss, is that the perturbative unitarization of modes $O(Q_S)$ prevents their diffusion into the infrared {\it \`a la} BFKL. Another consequence of gluon saturation is that it restores the separation of short and long distance modes characteristic of the OPE: The long-distance operators however are not the parton distribution functions extracted in Bjorken asymptotics but instead many-body correlators of lightlike Wilson lines that obey nonlinear extensions of the BFKL RG equation. 

As we stated earlier, the CGC is the semi-classical EFT describing the nonperturbative physics of gluon saturation. Because it is applicable in Regge asymptotics, its construction is natural in the infinite momentum frame (IMF) where the hadron lightcone momentum $P^+\rightarrow \infty$. The wee parton modes of interest are designated those with momenta $k^+ < \Lambda^+$, while the static hard lightcone modes occupy $\Lambda^+ < k^+\leq P^+$, where $\Lambda^+$ is a scale chosen such that $\alpha_S \ln(\Lambda^+/k^+)\ll 1$. It is further important to assume as well that the transverse momenta of both the wee and hard modes are $k_\perp \approx \mu$, where $\mu\gg \Lambda_{\rm QCD}$. 

The ideal hadron that realizes this IMF construction is an ultrarelativistic nucleus with atomic number $A\gg 1$~\cite{McLerran:1993ka,McLerran:1993ni}. Wee parton modes corresponding to a probe with $x\ll A^{-1/3}$ (where $x=k^+/P^+$) and $\Lambda_{\rm QCD} \ll k_\perp \approx \mu$ coherently resolve a large amount of color charge from the nucleons distributed within the $\sim 2 A^{1/3}$ diameter ($2 R_A$) of the large nucleus.\footnote{This occurs when the coherence length of the probe (the Ioffe length~\cite{Ioffe:1969kf}) 
$l_{\rm coh.}\sim 1/2m_N x \gg 2 A^{1/3}$, where the latter is the nuclear diameter in units of Fermi.} This wee probe ``sees" the fast moving modes of the nucleus as a ``shockwave" of overoccupied glue~\cite{McLerran:1994vd,Ayala:1995hx,Ayala:1995kg,Balitsky:1995ub}. The typical color charge representations are therefore classical, and in a random walk picture, are Gaussian distributed, with variance (the squared color charge per unit area) $\mu^2 \sim A^{1/3}$ fm$^{-2}$~\cite{McLerran:1993ka,McLerran:1993ni,Jeon:2004rk}. In the {\it gedankan} $A\gg 1$ limit, $\mu^2 \gg \Lambda_{\rm QCD}^2$; it therefore self-consistently fulfills the requirement that $k_\perp \approx\mu$ modes are semi-hard. Within the specified kinematic framework, the degrees of freedom are therefore static lightcone color charges and dynamical gauge fields, separated by the scale $\Lambda^+$.

Following these considerations, one can write down the CGC EFT generating functional: 
\begin{equation}
\label{eq:CGC-path-integral}
{\cal Z}[j] = \int [{\cal D}\rho] F_{\Lambda^+}[\rho]\Bigg[\frac{\int^{\Lambda^+}[{\cal D}A]e^{iS[A,\rho]+\int d^4 x \,j\cdot A}}{\int^{\Lambda^+}[{\cal D}A]e^{iS[A,\rho]}}\Bigg] \,.
\end{equation}
Here $j$ is an external current, $S[A,\rho]= S_{YM}[A]+S_{\rm coupling}[A,\rho]$ is the sum of the Yang-Mills action\footnote{Fermionic ``sea quark" degreees of freedom can be included systematically in this framework but for simplicity will be omitted in this discussion.} for the dynamical wee gauge field modes and $S_{\rm coupling}[A,\rho]$ represents a gauge-invariant coupling of these wee modes to the static classical color charge densities $\rho$ representing the modes with $k^+ > \Lambda^+$. 
Further, $F[\rho]$ is a gauge-invariant weight functional denoting the distribution of classical color charge densities. While as noted it is natural for the CGC weight functional to be Gaussian distributed in $\rho$ at the scale $\Lambda^+$ in a large nucleus, $F[\rho]$ is in general an arbitrary weight functional that captures the nonperturbative dynamics of the large $x$ modes. Convenient gauges for our weak coupling computations of high energy scattering are the lightcone gauge ($A^+=0$) or other axial gauges.  

Note that the structure of the path integral is such that one first performs the path integral over the gauge fields for each given distribution of color charge densities drawn from $F[\rho]$, before performing the average over the color charge densities. This implicitly assumes a large separation in the temporal (in $x^+$) evolution of these modes. This structure is very much that of a spin glass -- see for instance \cite{Gross:1984} - hence, the ``glass" in Color Glass. The condensate refers to the large occupancy of modes that make up $\rho$. A better conceptual understanding of both the glass and the condensate terminology, and the implications thereof, will be central to this work.

In the EFT construction we have outlined,  $\Lambda^+$ is an arbitrary scale, albeit within a specified kinematic range. 
The requirement that physical observables not depend on this scale is precisely what leads to a renormalization group evolution equation specifying how a given initial profile for $F[\rho]$ evolves with the change of scale $\Lambda^+$, or equivalently, the rapidity. This  JIMWLK RG equation~\cite{Jalilian-Marian:1997ubg,Jalilian-Marian:1998tzv,Iancu:2000hn,Iancu:2001ad} can be  reexpressed as equations (the Balitsky-JIMWLK hierarchy) for $n$-point lightlike Wilson line correlators in the shockwave background. For large $N_c$, and $A\gg 1$, the simplest 2-point Wilson line ``dipole correlator" (describing the cross-section in the shockwave background) obeys the closed form nonlinear Balitsky-Kovchegov (BK) equation~\cite{Balitsky:1995ub,Kovchegov:1999ua,Kovchegov:1999yj}. In a low density approximation corresponding to $g^2\mu\sim Q_S \ll k_\perp$, this equation reduces to the BFKL equation. The BK equation provides the simplest generalization of BFKL dynamics  incorporating the physics of gluon saturation. Its solution demonstrates the emergence of a dynamical saturation scale $Q_S(x)$ as an infrared fixed point of the equation and describes how this emergent scale solves the problem of infrared diffusion in BFKL dynamics. 

We will discuss the CGC EFT in more detail shortly. Computations in this framework are quite sophisticated just as in the BFKL case. The BK and JIMWLK RG equations have now been extended to NLLx accuracy. Process-specific ``impact factors" corresponding to higher order contributions (not accompanied by logarithms in $x$) have also been computed to next-to-leading order accuracy for a range of final states. Further, this framework is very versatile and has been applied to study a wide range of phenomena, from DIS to the early stages of heavy-ion collisions leading to the formation of the quark-gluon plasma~\cite{Berges:2020fwq}. 

While it seems likely that the CGC captures key features of gluon saturation, it is far from clear that it is the most general EFT consistent with the symmetries of the underlying theory. In particular, the word ``condensate" used to describe the overoccupied gauge field configurations suggests the breaking of global symmetries. While implicit in the framework, there is very little discussion in the literature on the symmetries that are broken, and what the consequences are. A discussion of symmetries is also relevant for the ``glass" feature of the CGC. The glass analogy is meaningful since the time scales for the dynamics of the ``impurities" (the random color charge densities) are much longer than the time scales of the dynamical wee parton modes; indeed, the former is taken to be static on the time scales of the latter. The breaking of symmetries however induces a dynamical time scale for the decay of the condensate and the vanishing of the glassy behavior, which qualitatively alters the CGC EFT on these time scales.\footnote{The time scales under discussion are qualitatively different from kinematic ``sub-eikonal" time scales that are typically much longer.} In other words, the semi-classical saturated gluon lumps are metastable as might be expected on general grounds~\cite{Coleman:1977hd}, with important qualitative and quantitative ramifications for high energy QCD. 

In this work, we will discuss the symmetries that are broken by the formation of  saturated gluon lumps in Regge asymptotics and their consequences. We stress at the outset that local gauge invariance of $SU(N_c)$ is always satisfied. 
The symmetries that are broken are those of Poincar\'{e} invariance and a global subgroup of $SU(N_c)$. The breaking of global symmetries (in particular the breaking of Poincar\'{e} invariance by the condensate) was discussed qualitatively previously~\cite{Dvali:2021ooc}. We will here discuss these more systematically and in greater detail. 

The global $SU(N_c)$ symmetry corresponds to the coset group $G=\G/\G_*$ of $SU(N_c)$ gauge transformations $g(x)$ (satisfying $A_\mu\rightarrow g^{-1}A_\mu g + g^{-1}\partial_\mu g$) where $\G$ and $\G_*$ differ, respectively, by the set of transformations $g(x)\rightarrow {\rm constant}$  and $g(x)\rightarrow 1$ for  $|\vec{x}|\rightarrow \infty$. For the more general case of $\G$, this constant is arbitrary and not necessarily unity.  The formation of the CGC breaks the global $SU(N_c)$ group $G\rightarrow \underbrace{U(1) \times U(1) \cdots \times U(1)}_{N_c - 1 \text{ times}}$, which is the residual symmetry group of the classical condensate field $\rho_{{\rm diag},s}$, where $s$ specifies the $N_c-1$ generators of the Cartan subalgebra $H$. 
In other words, $\rho_{{\rm diag},s}$ represents the diagonal elements of the color charge density $\rho^a$ we introduced previously. The broken generators $G/H$ of global $SU(N_c)$ are therefore Goldstone fields $U\in G/H$ representing excitations of the condensate whose dynamics is characterized by a Goldstone scale $f_G$ that appears in the kinetic term in the action for the fields $U$. This Goldstone scale is a new scale that appears in our generalization of the CGC EFT in addition to the previously discussed saturation scale $Q_S$. 
As we will show, the existence of this additional scale is what ensures that the EFT for saturated glue is a color superfluid, in contrast to the color glass wherein the dynamics of the Goldstone modes is frozen. 

In addition to the action for the Goldstone fields, an important element in our generalization of the CGC EFT is the coupling of the condensate to the colored Goldstones and to collinear gluon fields that are emitted on longer time scales relative to the reggeized gluons we discussed previously. In the CGC EFT, as we will discuss, the coupling of the color charge density to the wee gauge fields is expressed through the logarithm of a path ordered Wilson line. Since we now have additional Goldstone fields as well as collinear gauge fields in the generalized EFT, we will instead express the Wilson line in the  path integral using a coadjoint orbit action, the so-called Kostant-Kirillov-Souriau (KKS) action. 
This allows us to rewrite the term involving the Wilson line
in a simpler way  where path-ordering is eliminated when replaced by a functional integral over the colored Goldstone fields. In addition, the KKS action allows us to describe the 
depletion of the condensate by the emission of collinear gluons by introducing a color 
current for the gluon fields. Much of the emission of collinear gluons will be in the fragmentation region of the nucleus (rapidities $\ln(P^+/\Lambda^+)$) which are far separated from the rapidity of the probe ($\ln(s/Q^2)$). However this depends on the kinematics of the problem of interest. The radiation on long time scales can also occur at rapidities far from the fragmentation region 
 or not occur at all, resulting in a rapidity gap, as is the case in coherent diffractive scattering. This discussion of the decay of the condensate into collinear and wee modes can likely be formalized employing the methods of soft collinear effect theory (SCET)~\cite{Rothstein:2016bsq}.

Our resulting generalized EFT for gluon saturation has a rich structure that includes the frozen color charge condensate vacuum represented by $\rho_{\rm diag}$, the Goldstone fields $U$ which describe the fluctuations around this vacuum, and the collinear gluon fields ${\cal A}$ that they decay into. Together, the dynamics they describe is that of a color superfluid, not a static condensate, that flows on time scales that are parametrically of the order of the Goldstone dynamics. 
This can be estimated to be  $\tau_G\sim 1/{\cal N}\,\sigma$, where the cross-section for their scattering is $\sigma\sim 1/f_G^2$, and the density of condensate modes ${\cal N}\sim Q_S^3$, giving $\tau_G = f_G^2/Q_S^3$. In \cite{Dvali:2021ooc}, taking into account the additional breaking of translational invariance by the condensate, it was estimated that $f_G^2 = N\, Q_S^2$, where $N=O(1/\alpha_S)$ is the occupancy. Hence 
$\tau_G\sim 1/(\alpha_S Q_S)$, which is a large time scale in Regge asymptotics.\footnote{Note that for $Q_S\gg \Lambda_{\rm QCD}$, this time scale is much shorter than hadronization time scales since $\alpha_S\equiv \alpha_S(Q_S)$.} 

The Goldstone superfluid dynamics which is not important on time scales $\tau\sim 1/Q_S$ will be crucially important on this longer time scale. One should expect for instance that an IMF  probe containing modes $k^- \sim\alpha_S Q_S$ will be sensitive to excitations of the condensate. Similarly, in heavy-ion collisions, where two lumps of saturated glue scatter off each other, the corresponding characteristic lifetime\footnote{This is estimated in \cite{Baier:2000sb} employing kinetic theory to describe the scattering of the Goldstone excitations for occupancies $N > 1$. } describing the dynamics of excitations of the color superfluid is $\tau_G \sim \alpha_S^{-3/2} (Q_S)^{-1}$, with the additional $\alpha_S^{-1/2}$ factor resulting from the dominant expansion of the superfluid along the beam direction in the laboratory frame~\cite{Baier:2000sb}.



The outline of the paper is as follows. 
We begin by reviewing the CGC EFT in section~\ref{sec:CGC-EFT}. This is obtained by adapting and generalizing the Wong action describing the dynamics of color worldlines in background nonabelian gauge fields. The coupling of the wee gauge fields to the color charges is described by the logarithm of the Wilson line that captures their color precession in this background. 

In section~\ref{sec:Goldstone}, we provide a consistent treatment of the color charged states corresponding to saturated gluon configurations and discuss the consequences. We begin in Sec.~\ref{sec:Cold-cloud}
by first considering how one constructs color charged states by first imposing Gauss' law on such states to ensure that they are gauge-invariant. They can however transform under a Noether symmetry corresponding to global $SU(N_c)$ transformations that conserve color charge. Therefore, as discussed in Sec.~\ref{sec:Condensate}, to define saturated states with fixed large color charge densities, one needs to break this global symmetry to that of the Cartan subalgebra, resulting in the formation of the color condensate. The corresponding Goldstone effective action characterized by the novel scale $f_G$ is discussed in Sec.~\ref{sec:Goldstone-Kinetic-term}. 

Section~\ref{sec:Goldstone-dynamics} describes the dynamics of the condensate via its coupling to the Goldstone and collinear gauge fields. In section~\ref{sec:KKS}, we introduce the KKS action and the representation of Wilson lines in terms of this action. In particular, we construct the coarse-grained version of this action relevant to our discussion. We further discuss the application of the KKS action to discuss the coupling of the condensate to the Goldstone modes in Sec,~\ref{sec:KKS-Goldstone}, and to collinear gluons in Sec.~\ref{sec:Goldstone-decay}. Current conservation ensures that the Goldstone+collinear-glue dynamics will induce nontrivial currents in the condensate, leading to its depletion on time scales $\tau \geq \tau_G$.  We put together all the elements of the Goldstone effective action in Sec.~\ref{sec:Goldstone-final}. In order to do so, we will need to boost the Goldstone action of Sec.~\ref{sec:Goldstone-Kinetic-term}  constructed in the comoving frame of the condensate to the laboratory frame. Since the medium comprising the condensate breaks Poincar\'{e} invariance, this exercise is similar to computing the refractive index of a probe within a fluid in motion. 

As mentioned earlier, the effective action constructed in Sec.~\ref{sec:Goldstone-final} can be interpreted as that of a color superfluid. We discuss in section~\ref{sec:Superfluid} 
the corresponding ``Madelung" equations that describe the flow generated by the ``quantum" pressure of  the condensate. As in other superfluids, the color superfluid can develop a normal fluid component. In section~\ref{sec:QV-CGC}, we briefly summarize our construction of the color superfluid EFT that accounts for the Goldstone dynamics missing in the CGC EFT. We discuss further the consequences of their dynamics for experiments at current and future colliders.  

The appendices provide some of the details of the computations. In Appendix A, we discuss the quantization of the KKS action which results in the path integral for the Wilson line.  In Appendix B, we detail the transformation of Goldstone dynamics in the comoving frame to the laboratory frame. We also derive the expansion of the Goldstone action that generate the interactions of the colored Goldstone fields with each other and with collinear gluons.

\section{Review of Worldline formulation of Color Glass Condensate EFT}
\label{sec:CGC-EFT}
Here we will revisit the derivation of the CGC EFT in a worldline framework following the work of Jalilian-Marian, Jeon and Venugopalan~\cite{Jalilian-Marian:2000pwi}. As we
will see, this setup lends itself to the generalization we will discuss in the next section. Consider Wong's equations for classical color charged particles (worldlines) in a slowly varying background Yang-Mills gauge field. The former represent the small wavelength ($\sim 1/P^+$) large $x$ lightfront modes and the latter long wavelength ($\sim 1/k^+$) small $x$ modes where $k^+/P^+ \ll 1$. The Wong action \cite{Wong:1970fu} for such a system can be expressed as the gauge-invariant action
\begin{equation}
S_{\rm Wong} = -\frac{1}{4} \int F_{\mu\nu}^a F_a^{\mu\nu} - \sum_{j=1}^N \int d\tau \,m_0^j \sqrt{v_\mu^j v_j^\mu} + \frac{i}{N_c} \sum_{j=1}^{N} {\rm Tr} \{ Q_j \ln W_j\}\,.
\label{Wong}
\end{equation}
Here $N$ is the number of classically charged particles with  momenta greater than a scale  $\Lambda^+$ with color charges $Q_j^a$, and $v_\mu^j$ denotes the four-velocities of the particles. 
The color charges couple to the gauge fields through the path ordered Wilson line
\begin{equation}
W_j = {\mathcal P}\exp\left(ig \int_{-\infty}^\infty  d\tau\, v_j^\mu A_\mu^a(x_j^\mu(\tau))T_a^j\right)\,.
\label{eq:Wilson-1}
\end{equation}
It is important to note that the classical color charges $Q_j^a$ here are not classical in the sense of belonging to a high dimensional representation of the algebra. As discussed at length in \cite{Mueller:2019gjj}, they can be derived from the one loop effective action in QCD in a classical statistical approximation to the dynamics of Grassmann variables $\lambda_c^j$, whose bilinears define 
 $Q_j^a = {\lambda_c^j}^\dagger t_{cd}^a \lambda_d^j$, with the latter satisfying extended classical Dirac brackets $\left\{Q_j^a,Q_j^b\right\} = if_{abc} \,Q_j^c$, realizing the $SU(N_c)$ algebra. We will discuss an alternative to this Grassmann algebra formulation later. 
 
 For now, in connecting the Wong action to the  CGC EFT, one further assumes that $v^\mu = \delta^{\mu +}$ since the short wavelength modes are lightlike when $P^+\rightarrow \infty$. Further, because the Wong charges are static on the lightcone, the kinetic term in the Wong action can be dropped.  By ``static" in this formalism, we mean that the dynamics of these charges occurs on 
 times scales  $x^+ \sim 1/\Lambda^- = \Lambda^+/\Lambda_\perp^2$. For truly lightlike modes, assuming $\Lambda_\perp \ll P^+$, and $\Lambda^+ \sim P^+$, this is a very long time scale, which justifies their being treated as stochastic sources.  This is a point we will revisit because we will argue that the effective time scale for the color charge dynamics to set in is actually much shorter than $\Lambda^+/\Lambda_\perp^2$.  
 
 Dropping the kinetic term, the focus is then on the term coupling the static charges to the fields. A further approximation in the CGC EFT is the coarse graining of the $N$ color charges to define 
\begin{equation}
 \rho^a(x_\perp) = \sum_{j=1}^N \delta(x_\perp - x_\perp^j)\, Q_j^a \,.
 \label{CGC2}
\end{equation}
This goes hand-in-hand with assuming that the Wong charges are localized at $x_j^-=1/\Lambda^+ \rightarrow 0$. Their dynamics is entirely in the transverse plane.  More generally however in the RG enhanced CGC, this constraint is reinterpreted as $\rho_Y^a(x_\perp) = \rho^a(x^-,x_\perp)$, where $x^-$ corresponds to 
the averaged position of the color charges  on distance scales $< 1/\Lambda^+$, since $\Lambda^+$ is the sliding scale of the RG.

With these assumptions, the Wong action reduces to the CGC action
\begin{equation}
S_{\rm CGC} = -\frac{1}{4} \int F_{\mu\nu}^a F_a^{\mu\nu}  + \frac{i}{N_c} \int d^2 x_\perp\, {\rm Tr} \{ \rho(x_\perp) \ln W\}\,.
\label{CGC3}
\end{equation}
Because there is only one nonzero component of $v^\mu$, Eq.~(\ref{eq:Wilson-1}) simplifies here to 
\begin{equation}
 W = {\mathcal P}\exp\left(ig \int_{-\infty}^\infty  dx^+\, A_a^-(x^+,0,x_\perp)T_a \right)\,.
 \label{CGC4}
\end{equation}
 Defining 
\begin{equation}
 \ln W = -\sum_{k=1}^{\infty} \frac{1}{k} (1-W^k)\,,
 \label{CGC5}
\end{equation}
one can check from the equations of motion that the current density $J^+$ (and hence the corrresponding color charge $Q$) satisfies Wong's equations (with a specific boundary condition at $x^+=-\infty$), and is an element of the Lie algebra. More nontrivially, $\ln W$ itself is an element of the algebra~\cite{Lam:1998tw}.  It was shown further in \cite{Jalilian-Marian:2000pwi} that the action in Eq.~(\ref{CGC3}) generates the LLx BFKL equation.\footnote{The $\ln W$ action here differs from the action in 
\cite{Jalilian-Marian:1997qno} where $\ln W\rightarrow W$, by a cubic ``reggeon-reggeon-gluon" coupling term present in the latter. This term present in the latter action vanishes in the static recoil limit resulting in identical expressions in the two actions to $O(A^3)$ in the expansion needed to derive the BFKL equation. The $\ln W$ action therefore (unlike the $W$ action) reproduces BFKL without assuming recoil-less sources.} 

For a large number of frozen color charges $N$, the color charge density $\rho^a(x_\perp)$ in Eq.~\eqref{CGC2} corresponds to color charges in a higher dimension representation 
$Q$, whose elements commute: $[Q^a,Q^b]=\frac{\hbar}{N}\rightarrow 0$~\cite{McLerran:1993ka}. Since this will be highly relevant to our discussion in the rest of the paper, we will briefly summarize the detailed derivation by Jeon and Venugopalan~\cite{Jeon:2004rk}. For simplicity and clarity, we will restrict ourselves to the $SU(3)$ case; 
the  discussion is valid for arbitrary $N_c$.

An arbitrary $SU(3)$ representation can be labeled by the tensor indices $(m,n)$ where, in the Young tableaux representation, the upper tensorial index $m$ corresponds  to the difference in the number of boxes in the top row minus those in second row, and likewise, the lower index $n$ is the difference in the number of boxes in the second row to those in the bottom row. The $(m,n)$ space of a given representation is spanned by Cartan weights corresponding to the $N_c-1$ diagonal generators of the Cartan subalgebra. 
The phase space corresponding to the color charge $Q_j$ in Eq.~\eqref{CGC2} for an arbitrary large representation can be expressed as 
\begin{equation}
d^8 Q_j = d\phi_1 d\phi_2 d\phi_3 d\pi_1 d\pi_2 d\pi_3 dm dn \left( mn(m+n)\frac{\sqrt{3}}{48} \right)\,,
\label{CGC6}
\end{equation}
where $d\phi_i$ and $d\pi_i$ ($i=1,\cdots 3$) are the canonically conjugate Darboux variables spanning the symplectic volume (co-adjoint orbit) of the group~\cite{Alekseev:1988vx,Johnson:1988qm,Kelly:1994ig}. 
A key point noted in \cite{Jeon:2004rk} is that {\it if, and only if, the integrands in the phase space integral depend on $m$ and $n$ alone}, the volume integral over the Darboux variables can be performed exactly,  simplifying  
\begin{equation}
\label{eq:phase-space-su3}
\int d^8 Q_j = \frac{(2\pi)^3}{32 \sqrt{3}}\int dm dn\, m^2 n^2 (m+n)^2 \,.
\end{equation}

On the other hand, \cite{Jeon:2004rk} showed that multiplying the $N$ frozen charges 
results in the multiplicity
\begin{equation}
\label{eq:multsu3}
    M_{m,n}^{(N)} \approx  \frac{27 mn(m+n)}{N^3}\frac{3^{\frac{3}{2}+N}}{2\pi N}
    \exp\left(-3\frac{D_2^{m,n}}{N}\right)\left(1+3\frac{D_3^{m,n}}{N^2}\right)\,,
\end{equation}
where $D_2^{m,n}$ and $D_3^{m,n}$ are, respectively, the quadratic and cubic Casimirs for a given $SU(3)$ representation. We have assumed here that $m$ and $n$ are large, as befitting such a high dimensional representation. This multiplicity, multiplied by the dimension 
$d_{mn}\approx mn(m+n)/2$, of the representation, is nothing but the probability to find the system of $N$ frozen quarks in a given $(m,n)$ representation, and as such can be normalized appropriately:
\begin{equation}
\label{eq:probsu3}
    1= \frac{27\sqrt{3}}{4\pi N^4}\int dm dn\, m^2 n^2 (m+n)^2\,\exp\left(-3\frac{D_2^{m,n}}{N}\right)\left(1+3\frac{D_3^{m,n}}{N^2}\right)\,.
\end{equation}
Note that the phase space measure in this expression is identical to Eq.~\eqref{eq:phase-space-su3}, where~\cite{Jeon:2004rk,Jeon:2005cf}
\begin{equation}
    Q_j^a\rightarrow Q^a = \int d^2 {\bf x}_\perp \rho^a({\bf x}_\perp) \,,
\end{equation}
with the area under the integral being the coarse grained area containing the $N\sim A^{1/3}\gg 1$ large $x$ frozen color charges resolved by the small $x$ external probe.\footnote{Since these frozen charges typically have momenta $\Lambda^+$, $\Lambda_\perp$, $N$ here also corresponds to the phase space volume.}

It is thus in this manner that one obtains the {\it classical path integral} 
 in Eq.~\eqref{eq:CGC-path-integral}, with 
\begin{equation}
\label{eq:Gaussian}
    F[\rho] = \int d^2 {\bf x}_\perp\left( \frac{\rho^a\rho^a}{2\mu^2} -\frac{d_{abc}\rho^a\rho^b\rho^c}{\kappa}\right)\,.
\end{equation}
Here $\mu^2 = g^2 A/2\pi R_A^2$ is the color charge squared per unit area and $\kappa = g^3 A^2/\pi^2 R_A^4$ where we recall that $R_A$ is the nuclear radius.\footnote{For $\rho^a\sim \mu$, the cubic ``odderon" term is suppressed relative to the leading quadratic pomeron term by $A^{-1/6}$ in the large $A$ limit, as well as a further suppression in $N_c$.} In this case, the squared saturation scale $Q_S^2=\alpha_S N_c\mu^2$. This path integral is very suggestive that the color charge densities $\rho^a$ can be viewed as sources coupling to so-called reggeon fields (the fields corresponding to 
$\ln W$) which, as first observed in \cite{Jalilian-Marian:2000pwi}, are elements of the Lie algebra.\footnote{A complementary discussion and generalization of these ideas was subsequently given in the work of Caron-Huot~\cite{Caron-Huot:2013fea}.} Indeed, the weights in the above expression can be interpreted as those responsible for so-called color singlet pomeron and odderon excitations of reggeon field theory~\cite{Jeon:2005cf}. 

In the JIMWLK RG framework developed in \cite{Jalilian-Marian:1996mkd,Jalilian-Marian:1997qno,Jalilian-Marian:1997ubg,Iancu:2000hn,Iancu:2001ad,Ferreiro:2001qy}, $F[\rho]$ is treated as a more general nonperturbative distribution which is not restricted to the expression in Eq.~\eqref{eq:Gaussian}. The saturation scale is a dynamical quantity $Q_S\rightarrow Q_S(x)=Q_0 (1/x)^\lambda$, where $\lambda$ is the anomalous dimension for the evolution and $Q_0$ is a nonperturbative scale $Q_0\sim  A^{1/6}\Lambda_{\rm QCD}$. The JIMWLK framework is nevertheless a semi-classical framework which implicitly assumes the dominance of semi-classical lumpy color charge configurations whose distribution is governed by a stochastic path integral over classical color charges. 

Indeed, for each configuration of color charge densities, one can compute the saddle point of this path integral. This corresponds to the solution of the Yang-Mills equations 
\begin{equation}
\label{eq:YM-shockwave}
    D_\mu F^{\mu\nu,a} = J^{\nu,a}\,\,\,\, {\rm with}\,\,\,\, J^{\nu,a} = \rho^a({\bf x}_\perp)\delta(x^-)\delta^{\nu +}\,,
\end{equation}
with the other components of the lightcone current suppressed by powers of $1/\Lambda^+$. Exactly analogously to the QED case, there is a very simple static (independent of $x^+$) solution in the nonabelian case in Lorenz gauge ($\partial_\mu A^\mu=0$):
\begin{equation}
\label{eq:YM-Lorentz}
    A^{+,a} = -\frac{1}{\nabla_\perp^2}\rho^a \,\,;\,\, A^-=A^i=0\,,
\end{equation}
which corresponds to field strength $F^{ij}=0$, where here $i,j$ represent the two transverse coordinates. An alternative solution also satisfying this condition is 
\begin{equation}
\label{eq:classical}
    A^+=A^-=0\,\,;\,\, A^i = -\frac{1}{ig}V\partial^iV^\dagger\,\,{\rm with }\,\, V= {\mathcal P}\exp\left(-ig\int_{-\infty}^{x^-} dz^- \frac{\rho(x_\perp,z^-)}{\nabla_\perp^2}\right)\,.
\end{equation}
Here the path ordering is with respect to the smearing of the charge density in $x^-$. These ``shockwave"  classical configurations are characterized by singular field strengths on the transverse 2-D transverse plane of the nucleus and by vanishing field strength outside, corresponding to two distinct pure gauges for $x^-\lessgtr 0$. A plane wave color charge in the fundamental representation propagating through the pure gauge vacuum will experience a instantaneous rotation of its color charge by $V({\bf x}_\perp)$ at a transverse position ${\bf x}_\perp$ following its interaction with the shockwave at $x^-=0$; likewise, a color charge in the antifundamental representation will experience the color rotation $V^\dagger ({\bf y}_\perp)$ at the location ${\bf y}_\perp$.

One can solve the small fluctuation equations in the classical shockwave background to determine the shockwave propagators~\cite{McLerran:1994vd,Balitsky:1995ub,Ayala:1995hx,Ayala:1995kg}. These take a particularly simple form in the $A^-=0$ gauge~\cite{Hebecker:1998kv,McLerran:1998nk,Balitsky:2001mr}, where the gluon shockwave 
propagator in momentum space takes the form
\begin{equation}
\label{eq:shockwave-propa}
    D_{\mu\nu}^{ab}(p,q) = (2\pi)^4 \delta^{(4)}(p-q) D_{0,\mu\nu}^{ab}(p) 
    + D_{0,\mu\rho}^{ac}\,{\cal T}_{cd}^{\rho\sigma}(p,q)\,D_{0,\sigma\nu}^{db}(q)\,.
\end{equation}
Here $D_0$ is the free gluon propagator, and ${\mathcal T}$ is the shockwave effective vertex, which can be expressed as 
\begin{equation}
\label{eq:shockwave-vertex}
 {\cal T}_{ab}^{\mu\nu}(p,q)   = (2\pi) \delta(p^--q^-)(2p^-)g_{\mu\nu}\,{\rm sign}(p^-)
 \int d^2 {\bf z}_\perp e^{i({\bf p}-{\bf q})\cdot {\bf z}_\perp} V_{ab,\rm adj.}^{\rm sign(p^-)}({\bf z}_\perp)\,.
\end{equation}
An exactly analogous expression holds for fermion shockwave propagators with $2p^-\rightarrow \gamma^-$ and $V_{\rm adj.}\rightarrow V$, where V is the path ordered exponential in Eq.~\eqref{eq:classical} in the fundamental representation, $V_{\rm adj.}$ being its adjoint counterpart. 

Eqs.~\eqref{eq:classical} and \eqref{eq:shockwave-propa} are sufficient to compute amplitudes in the CGC EFT. Explicit NLO Feynman diagram computations with these building blocks have been performed for several DIS final states~\cite{Balitsky:2007feb,Boussarie:2016bkq,Roy:2019hwr,Caucal:2023fsf,Taels:2022tza,Bergabo:2023wed,Fucilla:2023mkl,Fucilla:2022wcg}. As previously noted, these shockwave computations allow one to separate out systematically the process dependent NLO impact factors from the process independent LLx JIMWLK/BK RG evolution equations.
An arbitrary amplitude  computed in this manner will then depend on the correlators $\langle V({\bf x}_\perp)\cdots V_{\rm adj.}({\bf z}_\perp)\cdots V^\dagger({\bf y}_\perp)\rangle_{\rho}$, whose evolution in rapidity, as noted earlier, is determined by the Balitsky-JIMWLK hierarchy of these lightlike Wilson line correlators.  The shockwave formalism has further been employed to compute the NLLx JIMWLK/BK RG equations, which are the state-of-the art~\cite{Balitsky:2013fea,Kovner:2013ona,Lublinsky:2016meo}.\footnote{We note however recent developments towards computing NNLx evolution~\cite{Caron-Huot:2016tzz,Brunello:2025rhh}.} 
We will discuss later in Section~\ref{sec:QV-CGC} the  impact of the dynamics of the condensate on these shockwave computations. 

\section{Colored Goldstones and a new scale in the CGC EFT}
\label{sec:Goldstone}
Following our brief review of the CGC EFT formalism, we will now discuss the breaking of a global $SU(3)$ symmetry that accompanies the formation of the condensate. We will show that this has implications that go beyond what has been discussed previously. In particular, we shall identify a new Goldstone scale in the generalized EFT which controls the temporal dynamics of the condensate. 

Because the condensate carries color charge, we will begin by first going over how colored states are defined in a nonabelian gauge theory. Much of this material is known 
\cite{Nair:2005iw} and expressed in different ways in different contexts in the literature. Our emphasis will of course be on the color states in high energy scattering. We will then discuss how the presence of large semi-classical color charge representations that are characteristic of gluon saturation allows us to employ the language of symmetry breaking. In particular, we will argue that the global $SU(3)$ symmetry is broken to a residual $U(1)\times U(1)$ symmetry carried by the charges constituting the condensate. The broken generators of global $SU(3)$ color correspond to  colored Goldstones that transform in the coset space $SU(3)/U(1)\times U(1)$. 

We will next write down the kinetic term in the ``low energy EFT" describing the dynamics of these Goldstone excitations controlled by a Goldstone decay constant. In section~\ref{sec:Goldstone-dynamics}, we will discuss how the coupling of the condensate degrees of freedom to the reggeon fields in the CGC EFT is modified by the presence of these Goldstone degrees of freedom.

\subsection{Gauge-invariant construction of charged states: Cold gluon clouds}
\label{sec:Cold-cloud}
We start with the canonical quantization of Yang-Mills theory in the standard Hamiltonian formulation in the $A_0 = 0$ gauge. The Hamiltonian is then given by
$H = {1\over 2} \int_V (E^2 + B^2)$ where $E^a_i = \partial_0 A^a_i$ is the electric field and $B^a_i= {1\over 2}\epsilon_{ijk} (\partial_j A_k^a - \partial_k A^a_j 
+ f^{abc} A^b_j A^c_k)$ is the magnetic field.
We are using the notation $A_\mu = (-it_a)A^a_\mu$, where
$\{ t_a\}$ are a basis for the fundamental representation 
satisfying $\Tr (t_a t_b ) = {\textstyle{1\over 2}} \delta_{ab}$ and
$[t_a, t_b ] = i f_{abc} t_c$. The integration in the expression for the
Hamiltonian is over a spatial volume $V$; the limit
$V \rightarrow \infty$ can be taken at the end as usual.
We also have the equal-time canonical commutation rule
\begin{equation}
[E^a_i (x) , A^b_j (y)] = -i \,\delta^{(3)}(x-y)\,.
\label{gauss1}
\end{equation}

We can represent the states of this field theory
by wave functionals $\Psi$
of the gauge field $A^a_i$; for consistency with the commutation rule,
$E^a_i$ is the functional derivative $(-i\delta /\delta A^a_i)$ acting on $\Psi$.
Gauss law, which is the equation of motion for $A_0$,
is not a Hamiltonian equation of motion since it does not involve
time-derivatives in terms of the phase space variables $(A^a_i, E^a_i)$.
It is imposed as a condition selecting the physical states of the
theory which obey 
\begin{equation}
(\partial_i E_i^a + f^{abc} A^b_i E^c_i ) \Psi
= (D\cdot E)^a \, \Psi = 0\,.
\label{gauss2}
\end{equation}
As is true for all operator equations in field theory, Eq.~(\ref{gauss2}) must be understood as valid after smearing with suitable test functions.

While of no consequence for many calculations
in field theory, and therefore often omitted, the nature of the test function
is important for a proper definition of  charged states.
We define 
the smeared operator\footnote{We will see shortly that the smearing function is an angular variable, so we use $\theta^a$ in anticipation.}
\begin{equation}
{\cal C}_0 (\theta ) =  \int_V d^3x\, (\theta^a (D\cdot E)^a)\,\,\,{\rm with}\,\, \,{\cal C}_0 (\theta) \Psi = 0\,,
\label{gauss3a}
\end{equation}
where the latter equation is Gauss Law, valid for any $\theta^a$.
Thus if $\theta^a$ and $\theta'^a$ are two different choices for
the test function, then any physical state
$\Psi$ will obey both 
${\cal C}_0 (\theta) \Psi = 0$ and ${\cal C}_0 (\theta') \Psi = 0$. We clearly then have the consistency requirement
\begin{equation}
{\cal C}_0 (\theta) \Psi = {\cal C}_0 (\theta') \Psi = 0 ~~\Longrightarrow~~
[{\cal C}_0 (\theta), {\cal C}_0 (\theta')] \Psi = 0\,.
\label{gauss3b}
\end{equation}
By direct use of the canonical commutation rule in Eq.~(\ref{gauss1}), we can express the commutator as
\begin{equation}
[{\cal C}_0 (\theta ), {\cal C}_0 (\theta' ) ] = i {\cal C}_0 (\theta \times \theta') - i \oint_{\partial V}
(\theta \times \theta')^a E^a_i dS_i\,,
\label{gauss4}
\end{equation}
where $(\theta \times \theta')^a = f^{abc} \theta^b \theta'^c$.

While we expect ${\cal C}_0 (\theta \times \theta') \Psi$ to vanish
as well, the second term on the right hand side 
of Eq.~(\ref{gauss4}) can be problematic. For this term to vanish, we need 
the functions $\theta^a$ to vanish on the boundary $\partial V$.
In other words, to impose Gauss law consistently on the
wave functions, we must use test functions which vanish on
$\partial V$. Notice further that
\begin{equation}
e^{i {\cal C}_0(\theta)} e^{i {\cal C}_0(\theta')}  = e^{i {\cal C}_0(\theta) + i {\cal C}_0(\theta')  -{1\over 2} [{\cal C}_0(\theta), {\cal C}_0(\theta')] + \cdots} \,.
\label{gauss4a}
\end{equation}
Combining the unitary transformations generated by ${\cal C}_0 (\theta)$ to produce
$e^{i {\cal C}_0 (\alpha )}$ (where $\alpha$ is some combination of
$\theta$ and $\theta'$) will require that the commutators should
be proportional to ${\cal C}_0$. This again leads to the requirement
that $\theta \rightarrow 0 $ on $\partial V$.
Thus one can understand the subscript of ${\cal C}_0$ in Eq.~(\ref{gauss3a}) as requiring that $\theta$ vanishes on $\partial V$.

With $\theta \rightarrow 0$ on $\partial V$, we see that ${\cal C}_0 (\theta)$ generates an infinitesimal gauge transformation of $A^a_i$
by a group element $ g = e^{-i t_a \theta^a}$:
\begin{equation}
e^{i {\cal C}_0(\theta)} \, A\, e^{-i {\cal C}_0(\theta)} \approx 
A + D \theta 
 \approx g^{-1} A g + g^{-1} \partial g, \hskip .3in
 g = e^{-i t_a \theta^a} \approx 1 - i t_a \theta^a
 \label{gauss3}
 \end{equation}
 A further consequence is that 
 \begin{eqnarray}
 - i {\cal C}_0 (\theta )\Psi = i \int D\theta \cdot E \Psi
 = \int (D_i\theta)^a {\delta \over \delta A^a_i}  \Psi
 = \Psi (A+ D \theta) - \Psi (A)  
 = 0\,.
 \label{gauss4b}
 \end{eqnarray}
This confirms that physical states are gauge-invariant under gauge transformations
$A \rightarrow A^g$ where $g \rightarrow 1$ on $\partial V$
or as $\vert \vx \vert \rightarrow \infty$ if we take the limit of infinite volume.

Another useful observation is the following. If $\Psi$ is a physical state,
an operator ${\mathcal O}$ can act on it and generate a physical
state if 
\begin{equation}
{\cal C}_0 (\theta) \, {\mathcal O} \Psi =
[{\cal C}_0 (\theta), {\mathcal O}] \Psi = 0\,.
\label{gauss4c}
\end{equation}
Thus the statement that all physical observables must weakly commute with ${\cal C}_0(\theta)$ 
is equivalent to their commutator vanishing on physical states. 

We will now analyze the nature of the transformations
$A \rightarrow g^{-1} A g + g^{-1} \partial g$
where $g \neq 1$ on $\partial V$ or at spatial infinity.
We begin by defining the
operator
\begin{equation}
{\cal C}(\theta) = - \int (D_i\theta)^a E_i^a = 
\int d^3x\, \theta^a ( D\cdot E)^a - \oint \theta^a E^a\cdot dS\,,
\label{gauss6}
\end{equation}
which can be contrasted with the definition of ${\cal C}_0$ in Eq.~(\ref{gauss3a}).
The last expression in the above equation shows that this operator differs from Gauss law
by a surface term which is not necessarily zero.
It is easy to check by use of the canonical commutation rules that
\begin{equation}
[{\cal C}(\theta), {\cal C}(\theta') ] = i {\cal C}(\theta\times \theta')\,.
\label{gauss7}
\end{equation}
We emphasize that the closure of the algebra of ${\cal C}$'s in this result (unlike Eq.~(\ref{gauss4})) does not require us to discard any surface terms; it holds for $\theta \neq 0$ on $\partial V$.
Its closure under commutation shows that
we can define finite transformations by exponentiation of
${\cal C}(\theta)$. We can also combine such transformations
to produce new ones.

Since $(\theta \times \theta')^a$ will vanish on $\partial V$
if $\theta'^a$ vanishes on $\partial V$, the commutation rule
Eq.~(\ref{gauss7}) also shows that
\begin{equation}
[{\cal C}_0(\theta'), {\cal C}(\theta) ] = -i {\cal C}_0(\theta\times \theta')\,.
\label{gauss7a}
\end{equation}
Thus the commutator will vanish on physical states. Therefore in agreement with
Eq.~(\ref{gauss4c}), ${\cal C} (\theta)$ will generate
physical states from the physical state $\Psi$; in other words,  
${\cal C}(\theta)$ is an observable.

From Eq.~(\ref{gauss6}) we see that ${\cal C}(\theta)$, up to Gauss' law,
is characterized by the value of $\theta$ on $\partial V$.
The corresponding states created by the application of
${\cal C}(\theta)$ are therefore referred to as edge states.
In terms of the action on $A^a_i$, we also have
\begin{equation}
e^{i {\cal C}(\theta)} \, A\, e^{-i {\cal C}(\theta)} 
= g^{-1} A g + g^{-1} \partial g
\label{gauss7b}
\end{equation}
where, once again, $g \neq 1$ on $\partial V$.

Therefore in general, $\theta^a$ in $g = e^{-i t_a \theta^a}$ can be any function
on $\partial V$. If we consider the large
volume limit, $\theta^a$ can also have angular dependence
as we take $\vert \vx \vert \rightarrow \infty$.
Such transformations and edge states have been considered in \cite{Balachandran:1993tm,Balachandran:1994vi,Balachandran:2019bzg,Balachandran:2021sze}.
These transformations also correspond to the asymptotic symmetries of the theory. In the present context, it is useful to consider a subset
of such transformations where $\theta^a$ is a constant, independent of angles. 
In this case, we see that we can write for a physical state $\Psi$,
\begin{equation}
e^{i {\cal C} (\theta )} \, \Psi = e^{- i\theta^a_\infty Q^a} \Psi, \hskip .3in
Q^a = \oint_{\vert\vx\vert\rightarrow \infty} E^a\cdot dS 
\label{gauss8}
\end{equation}
The Gauss law piece in Eq.~(\ref{gauss6}) vanishes on $\Psi$ and
$\theta^a_\infty$ is the constant value of $\theta^a$
as $\vert \vx \vert \rightarrow \infty$.
This shows that the states transform under the action of
$e^{i {\cal C}(\theta)}$ as representations of the group $G$.
The group generators $t_a$ (in $g = e^{-i t_a \theta^a}$)
are given by the charge operators $Q^a$ in implementing
the transformation on
physical states $\Psi$. Eq.~(\ref{gauss8}) also shows that
these transformations act like a Noether symmetry, namely, 
a rigid or global symmetry, on the states $\Psi$.

To recapitulate, gauge transformations $g$ which tend to
the identity on $\partial V$ or as $\vert \vx \vert \rightarrow
\infty$ are the true gauge transformations of the theory
in the sense that they characterize the redundancy 
in using the gauge potentials. Such transformations are generated
by the Gauss law ${\cal C}_0 (\theta)$ which annihilates
physical states.
The transformations $g$, where $g$ is a constant element of $G$ that is 
not necessarily $1$ on $\partial V$,
act as a Noether symmetry. These states can be classified into
representations of $G$, and for nontrivial representations, are the charged states.

We will also comment briefly here about how matter fields
are included in this analysis.
Consider a matter field $\psi$ which transforms under the gauge transformation
as $\psi \rightarrow g \psi$, with $g$ in the appropriate representation.
This field is not acceptable for
constructing charge-carrying physical states as $\psi^\dagger \ket{0}$
since it is not gauge-invariant; it is not invariant even under transformations
which are the identity as
$\vert \vx \vert \rightarrow \infty$.
To obtain a physical state one must use
a composite field ${\tilde V} \psi$, where ${\tilde V} $ is an operator which cancels out the
variation of $\psi$ under a gauge transformation, namely, under the action of
$e^{i {\cal C}_0 (\theta)}$. In the case of QED, this was first discussed by Dirac \cite{Dirac:1955uv}, who defined the gauge-invariant operator
\begin{equation}
{\tilde V} \, \psi (x) = e^{ -i e \int_y  D(x, y) \nabla \cdot A (y) } \, \psi(x)\,,
\label{gauss10}
\end{equation}
where $D(x, y)$ here is the Coulomb Green's function obeying $-\nabla^2_x D(x, y) 
= \delta^{(3)}(x-y)$. More generally, ${\tilde V}$ is  related to infrared dressing factors, see 
Ref.~\cite{Weinberg:1995mt} and references therein, as well as \cite{Kulish:1970ut},
and \cite{He:2014cra,Campiglia:2015qka,Kapec:2015ena} for more recent work.\footnote{Asymptotic symmetries and infrared dressings of QED fields have been discussed in the worldline framework in \cite{Feal:2022iyn,Feal:2022ufw}.}
For dressing factors on charged fields in the nonabelian theory, see Ref.~\cite{Lavelle:1995ty}. These ``cold gluon clouds" can be understood further in the universal language of soft anomalous dimensions of Wilson line correlators~\cite{Korchemsky:1991zp,Gardi:2025auw}. In particular, in the high energy Regge limit, gluon reggeization, which is a building block of BFKL dynamics can be formulated in this language~\cite{Korchemskaya:1996je,Falcioni:2021buo}.

To complete our discussion of matter fields, we note that they modify the
definition of 
${\cal C}_0(\theta)$ and ${\cal C}(\theta)$.
With $J^{0a}$ as the charge density for the matter fields,
the modified operators are
\begin{eqnarray}
{\cal C}_0(\theta) &=& \int d^3x\, \left( \theta^a (D\cdot E)^a - \theta^a J^{0a} \right)\,.\nonumber\\
{\cal C}(\theta) &=& \int d^3x\, \left(- D_i \theta^a E^a_i  - \theta^a J^{0a} \right)\,.
\label{gauss11}
\end{eqnarray}

To summarize, we can formalize the statements about the role of ${\cal C}_0 (\theta)$ and
${\cal C}(\theta)$ by defining
two sets of transformations:
\begin{eqnarray}
\G_* &=& \bigl\{ {\rm Set~of~all~} g (x)=e^{i {\cal C}_0 (\theta )}: \mathbb{R}^3 \rightarrow SU(N_c) ~{\rm such ~that~} g(x) \rightarrow 1 ~{\rm as~} \vert \vx \vert \rightarrow \infty\bigr\}\,.
\nonumber\\
\G &=& \bigl\{ {\rm Set~of~all~} g (x)=e^{i {\cal C} (\theta )}: \mathbb{R}^3 \rightarrow SU(N_c) ~{\rm such ~that~} g(x) \rightarrow g_\infty ~{\rm as~} \vert \vx \vert \rightarrow \infty\bigr\}\,.
\label{gauss5}
\end{eqnarray}
Here $g_\infty$ is taken to be a constant element of $G$, not necessarily the identity. In other words $g_\infty = e^{-i t_a \theta_\infty} \in G$
as in Eq.~(\ref{gauss8}).
These two sets $\G$ and $\G_*$ differ by the boundary values $g_\infty$, so we can also
see that $\G /\G_* = G$. The exponentiation of the Gauss law, namely, $e^{i {\cal C}_0(\theta)}$
corresponds to the quantum version of transformations by elements
of $\G_*$ and likewise, $e^{i {\cal C}(\theta)}$ corresponds to the transformation by elements of $\G$. As mentioned before, $\G_*$
 is the true gauge group of the theory in the sense that
all wave functions are $\G_{*}$-invariant, 
including finite transformations.
The coset group $\G /\G_* = G$ acts as a global symmetry on the physical states. In a gauge theory, the true gauge symmetry (or more accurately, the  gauge redundancy) $\G_*$ is never broken even when there is spontaneous symmetry breaking. 
It is only $\G$, or more precisely $G = \G/\G_*$, which is broken.
\subsection{From cold gluon cloud to overoccupied condensate: Symmetry breaking and colored Goldstones}
\label{sec:Condensate}
We considered thus far the general problem of the construction of gauge-invariant color charged states. For the problem of interest--the dynamical formation of a saturated gluon state due the overoccupancy of gluons in Regge asymptotics--the  CGC EFT 
treats such states as semi-classical color charge distributions. We will discuss here the relevant symmetry structure for such overoccupied gluon configurations in light of our discussion in the previous subsection. 

Specifically, if we wish to construct the path integral in 
Eq.~(\ref{eq:CGC-path-integral}) from the QCD Hamiltonian, the coarse-grained intermediate state corresponding to a given large color charge representations must lie in the physical Hilbert space. In other words, such states must be annihilated by the Gauss law operator ${\cal C}_0$. However because 
the color charge density operator $J^{0a}$ is only gauge-covariant and not gauge-invariant, specifying physical states of fixed given large color charge is problematic. 

To clarify this further, consider a  physical state $\ket{\Psi}$ in the path integral construction for which the charge density has a large specific value\footnote{In the CGC case, this large value is $O(1/g)$ with quantum fluctuations around this value at most $O(1)$. For a generic ``cold gluon cloud" configuration, quantum fluctuations are large and $\rho^a$ is not the dominant configuration.}
given by a classical function $\rho^a=\bra{\Psi} J^{0a} \ket{\Psi}$.
But we can also write it as 
\begin{eqnarray}
\rho^a 
=\bra{\Psi}(e^{-i {\cal C}_0(\theta)}  J^{0a}\, e^{i {\cal C}_0(\theta)} )\ket{\Psi} 
 = \D^{ab}(\theta) \, 
\bra{\Psi}  J^{0b}  \ket{\Psi}
=\D^{ab}(\theta) \, \rho^b\,,
\label{gauss12}
\end{eqnarray}
where $\D^{ab}$ is a  matrix element of
$g= e^{ -i t_a \theta_a}$ in the adjoint representation and 
we have used the commutation rule $[{\cal C}_0(\theta), J^{0a}(x)]
= - i f^{cab} \theta^c J^{0b} (x)$.
We also used here the fact that because 
$\ket{\Psi}$ is a physical state, $e^{i {\cal C}_0(\theta)}  \ket{\Psi} = \ket{\Psi}$. From Eq.~(\ref{gauss12}) we see that if we specify a nonzero expectation value for $J^{0a}$, the state
cannot be taken to be invariant under the full ${\cal C}_0(\theta)$,
except for a subgroup of transformations
for which $\rho^a = \D^{ab} \rho^b$.
It would therefore be incompatible with the Gauss law constraint on 
physical states.

To see how this apparent problem with Gauss' law is resolved,
recall that a nonabelian charge density operator
$\hat{\rho} = J^{0a} t_a$ will transform under gauge transformations as\footnote{This is equivalent to the operator relation $J^{0a} \rightarrow \D^{ab} J^{0b}$ when written in the adjoint representation.}
\begin{equation}
{\hat\rho} \rightarrow U^{-1} {\hat \rho} U, \hskip .3in {\hat \rho} = {\hat \rho}^a t_a, ~~U\in {\cal G}_*\,.
\label{syn8}
\end{equation}
In Eq.~(\ref{syn8}), $U$ is $x$-dependent, 
so we can choose $U(x)$ pointwise to diagonalize ${\hat \rho}$.
Thus, up to a gauge transformation, a nonabelian charge density operator is specified
by just the diagonal color charge density operators which are the densities in the Cartan subalgebra.
This is also in agreement with the fact that the gauge-invariant information in the charge density is given by the Casimir-type invariants we can construct from $J^{0a}$;
these are traces of products of ${\hat\rho}$, which are all 
fixed by the diagonal values of ${\hat\rho}$.

In principle, $U$ can be eliminated in a suitable gauge.
However if we have already fixed a gauge, $U$ can no longer be eliminated while retaining the same gauge condition for $A_\mu$. The degrees of freedom in $U$
should be therefore be counted as physical degrees of freedom.
Being a specific matrix which diagonalizes ${\hat \rho}$,
the elements of $U$ carry information about ${\hat\rho}$.
Taking $U^{-1} {\hat \rho} U$ as diagonal, we can write
\begin{equation}
{\hat \rho} = U \, {\hat \rho}_{\rm diag} U^{-1}\,.
\label{syn9}
\end{equation}
Notice that if we make the transformation
$U \rightarrow U \, u$, where $u$ is an element of the
Cartan subgroup $H$ of $G$, ${\hat \rho}$ is unchanged. This implies that $U$ is defined
only up to this transformation. It belongs to the coset space $G/H$.

Thus the potential problem with Gauss law in the path integral construction is resolved by requiring that the color charge density of a given dynamically generated saturated gluon state is specified by the expectation value of the color charge density operator $\rho_{\rm diag}=\bra{\Psi}{\hat \rho}_{\rm diag}\ket{\Psi}$. Unlike ${\hat \rho} $, which  transforms as 
${\hat \rho} \rightarrow g {\hat \rho} g^{-1}$ under a
gauge transformation, ${\hat \rho}_{\rm diag}$ is gauge-invariant, if correspondingly, $U \rightarrow g U$. In particular,  ${\hat \rho}_{\rm diag}$ commutes with the Gauss law operator ${\cal C}_0$. Further, it is appropriate for the large charge semi-classical configurations in the CGC EFT that ${\hat \rho}_{\rm diag}$ in the path integral construction be replaced by its expectation value $\rho_{\rm diag}$.  

However because $\rho_{\rm diag}$ is not invariant under global transformations $G$,  it must spontaneously break this global 
symmetry for it to correspond to a unique gauge-invariant color charge density.\footnote{In other words, under the larger symmetry group ${\cal G}$ of $SU(N_c)$, ${\hat \rho}_{\rm diag}$ would transform as ${\hat \rho}_{\rm diag} \rightarrow e^{i{\cal C}(\theta)}{\hat \rho}_{\rm diag}e^{-i{\cal C}(\theta)}$. In that case, 
\begin{equation}
\rho_{\rm diag} = \bra{\Psi}{\hat \rho}_{\rm diag}\ket{\Psi} \equiv  \bra{\Psi}e^{i{\cal C}(\theta)}{\hat \rho}_{\rm diag}e^{-i{\cal C}(\theta)}\ket{\Psi}= \bra{\Psi}e^{i\theta^a_\infty Q^a}\,{\hat \rho}_{\rm diag}\,e^{- i\theta^a_\infty Q^a} \ket{\Psi}\,,
\end{equation}
where we used Eq.~(\ref{gauss8}). Since $[Q,{\hat \rho}_{\rm diag}]\neq 0$ except for the diagonal generators of $Q$, $\rho_{\rm diag}$ will not remain invariant under this asymptotic symmetry. Thus if $\rho_{\rm diag}$ represent a gauge-invariant condensate, it must break this global symmetry.
} The corresponding spontaneously broken state unambiguously defines the condensate in the Color Glass Condensate. It still possesses a $U(1)\times U(1)$ residual symmetry, transforming under the diagonal generators belonging to $H$. The six non-diagonal generators belonging to $G/H$ represent the symmetries broken by the ``CGC vacuum". 

Specifically, we may regard the parametrization 
in Eq.~(\ref{syn9}) describing small spacetime-dependent fluctuations $U$ around the diagonal value to be the approximate ``Goldstone-type" fields for our
problem.
Of the ${\rm dim\,}G$ degrees of freedom in
$\rho = \rho^a t_a$, ${\rm rank\,}G$ are included
as $\rho_{\rm diag}$, while $U$ accounts for 
$({\rm dim\,}G - {\rm rank\,}G)$ coset degrees of freedom. Thus Eq.~(\ref{syn9}) is a proper and complete parametrization of 
$\rho$. In summary, the parametrization in Eq.~(\ref{syn9}) allows for a 
gauge-invariant description of physical situations with a large fixed charge density $\rho_{\rm diag}$ containing  ${\rm rank\,}G$ degrees of freedom.

It is useful to compare this understanding with the description in section~\ref{sec:CGC-EFT}.
There the color charge was separated into a part which lies in the
Cartan subalgebra and quantified by large $m, n$, which correspond to $\rho_{\rm diag}$. 
The coset elements in $U$ are the other part, the 
Darboux coordinates in the integration measure
in Eq.~(\ref{CGC6}). The further simplifications, and the formulae given in
Eqs.~(\ref{eq:phase-space-su3}) and (\ref{eq:multsu3}), 
considered the approximate situation where the integrands are 
independent of the coset variables.
In other words, the dynamics of $U$ was not considered there and the entire charge density operator $\hat \rho$ was treated as a nondynamical degree of freedom. 
Our main point here is to show that the results of
section~\ref{sec:CGC-EFT} must be augmented by including the Goldstone degrees of freedom.\footnote{A piece of conventional wisdom regarding the CGC EFT is that the ``Gaussian" approximation for the weight functional $F[\rho]$ in Eq.~(\ref{eq:Gaussian}) is qualitatively modified by the RG evolution. (With each step in the evolution, an induced charge $\delta \rho$ is added to the color charge density at the previous step.)  However an alternative perspective is that the nondynamical $\rho_{\rm diag}$ distribution always remains Gaussian distributed but the quantum fluctuations accounted for by RG evolution have nontrivial dynamics (captured by $U$) in the symplectic phase space of the broken Darboux generators. This may explain why the RG evolved $F[\rho]$ can still be approximated by Gaussian-like distribution albeit with a nonlocal variance~\cite{Iancu:2001md}.} The frozen degrees of freedom are represented here by $\rho_{\rm diag}$ alone while the Goldstone dynamics of the $U$ fields makes 
$\hat \rho$ dynamical. In other words, this is fundamentally how its expectation value $\rho(x_\perp,x^-)\rightarrow \rho(x_\perp,x^-,x^+)$.

Strictly speaking, what we are describing is not spontaneous symmetry breaking.
This is because in principle we can have independent dynamics for
$\rho_{\rm diag}$ as well as for $U$ because there are operators
which can change $\rho_{\rm diag}$.
(In other words, it does not have to be ``frozen".)
A change in $U$ of the form $U \rightarrow U R$ for some
$R \in G$ can be compensated by a change of $\rho_{\rm diag}$,
which need not remain diagonal. 
However when the state has a large charge which is approximately fixed
and we only consider
small fluctuations of the charge, 
the language of spontaneous symmetry breaking becomes useful. This is the case in our physical construction of the CGC EFT, where the modes ($\Lambda^+ < k^+ \leq P^+$) have large occupancies $O(1/g)$, and are effectively classical. Further, as noted, these modes are quasi-static modes on the light cone. One thus arrives at a physical picture of the Goldstones as small frequency fluctuations around the ``frozen" $\rho_{\rm diag}$ color charge density. 

A simple analogy can help clarify these statements.
Consider a ferromagnet of finite volume $V$ or containing a finite ${\mathcal N}$ number of degrees of freedom. At low temperatures, the ground state $\ket{n}$ has nonzero magnetization
$\bra{n} M_i \ket{n} = C n_i$ along some direction specified by
the unit vector $n_i$.
The matrix element $\braket{n'}{n}$ which corresponds to a transition
which changes the orientation of all the molecular spins to obtain a 
reorientation of $\langle M_i \rangle$ is exponentially suppressed 
as $e^{- {\mathcal N} (\delta\theta)^2/8}$ for a small angular difference $\delta \theta$ between $n_i$ and $n'_i$ for
$\mathcal N$ spin-${1\over 2}$ molecular magnets. For ${\mathcal N} \rightarrow \infty$, this matrix element is zero and we have
spontaneous symmetry breaking. A local reorientation of a few 
molecular spins is not suppressed; these would give spin waves or the Goldstone modes. For finite but large ${\mathcal N}$, although we do not have spontaneous symmetry breaking in the strict mathematical sense,
we do have the same general features. The background magnetization is large with small local fluctuations, since the transition element
$\braket{n'}{n}$ is still very small. This is the sense in which we use the language of symmetry breaking in the CGC. 

\subsection{Kinetic term and decay constant for  Goldstone fields}
\label{sec:Goldstone-Kinetic-term}
It is straightforward to construct an EFT for the Goldstone modes $U$.
The effective action, as usual, should be given as a set of
terms made of monomials of fields and their derivatives
which are consistent with the symmetry structure and 
arranged as a series in terms of increasing number of derivatives.
We will start with a kinetic term involving the minimal number of derivatives.
The relevant symmetries for us, as noted previously, are the gauge transformation
on $U$ given by $U \rightarrow g U$ and 
the fact that $U$ belongs to $G/H$.
The latter requirement implies that the terms we use are invariant
under $U \rightarrow U u$, where $u$ is in the Cartan subgroup.
Gauge invariance requires that the derivatives should be covariant derivatives, gauged on the left of $U$. In other words, we define 
${\bar D}_\mu U = \partial_\mu U + A_\mu U$, $A_\mu = (-i t_a)A^a_\mu$ which, importantly, distinguishes it from the usual covariant derivative $D_\mu C = \partial _\mu C + g [A,C]$, for an arbitrary field $C$.

Consider first the set of matrices
which form an orthonormal basis for the
Lie algebra of $G$ in the fundamental
representation. We will split the set $\{ t_a\}$ into two: $h_s$, $s=1, 2, \cdots, r$ are the diagonal matrices (generators of $H$), where $r$ is the rank of the group ($r=2$ for $N_c=3$). 
The set $\{ E_A\} $, where $A$ takes ${\rm dim}\,G
- {\rm rank}\,G$ values, are the coset generators corresponding to the tangential
directions of $G/H$. 

In the Cartan-Weyl basis, the generators $E_A$ can be further divided 
into raising and lowering operators, $E_{\bf {\talpha}}$ and $E_{-{\bf \talpha}}$, respectively, where
${\bf \talpha}$ denotes a root. These satisfy
\begin{equation}
[h_s, E_{\bf \talpha} ] = \talpha_s \, E_{\bf \talpha}\,,
\hskip .3in
[h_s, E_{-\bf \talpha} ] = - \talpha_s \, E_{-\bf \talpha}\,.
\label{syn9a}
\end{equation}
As a result, we see that 
\begin{equation}
\Tr \left( E_{\pm \bf\talpha} (Uu)^{-1} {\DU}_\mu (Uu)\right)
= e^{\pm  i \talpha_s \theta_s } \Tr \left( E_{\pm \bf\talpha}
U^{-1} \DU_\mu U\right)\,,
\label{syn9b}
\end{equation}
where $u = e^{i h_s \theta_s}$ is an element of the Cartan subgroup
$H$. In arriving at the right hand side of
Eq.~(\ref{syn9b}), we used $\Tr( E_{\pm\bf\talpha} h_s ) = 0$, 
which ensures that the term proportional to $\partial_\mu \theta_s$ is zero.
This transformation rule shows that the product 
 $\sum_{\talpha} \Tr \left( E_{\bf\talpha}U^{-1} \DU_\mu U\right) \Tr \left( E_{-\bf\talpha}
U^{-1} \DU_\nu U\right)$ is invariant under gauge transformations
(because we use the covariant derivative)
and under $U \rightarrow U \, u$.

If these considerations applied to symmetry breaking by the vacuum state, Lorentz invariance would be required and we would need to contract the $\mu, \nu$ indices with the Minkowski metric.
However, in the present case, we are describing the condensate; 
therefore, $\mu, \nu$ should be contracted with
$\sigma^{\mu\nu}$ which is a tensor characteristic of the medium.
This tensor $\sigma^{\mu\nu}$ differs from $\eta^{\mu\nu}$ 
by a quantity that is analogous to the polarization tensor $\Pi^{\mu\nu}$
in media with electromagnetic interactions. In principle, 
$\sigma^{\mu\nu}$ can be calculated from the material properties of the condensate. But in the spirit of constructing an effective 
action (where the only dynamics of relevance is what the external probe is sensitive to), we will take it to be a given quantity.
The leading term of the effective action 
(with the minimal number of derivatives of $U$)
is thus given by\footnote{Just as in chiral perturbation theory, one can add higher derivative terms to the action. We will discuss these later.}
\begin{equation}
S_{\rm Goldstone} = -{1\over 2} f_G^2\int  \sigma^{\mu\nu} \sum_{\talpha} \Tr \left( E_{\bf\talpha}
U^{-1} \DU_\mu U\right) \Tr \left( E_{-\bf\talpha} U^{-1} \DU_\nu U\right)\,.
\label{syn9c}
\end{equation}
This action provides the kinetic term for the dynamics of $U$, 
consistent with the symmetry structure of gauge covariance,
invariance under $U \rightarrow U \, u$ and
the reduced Lorentz symmetry of the medium.
The dimensionful parameter $f_G$ is the analog of the
pion decay constant. It provides the time scale for the interactions of the low energy color fluctuations. 
In general, $\sigma^{\mu\nu}$ is a symmetric tensor with
distinct components reflecting the possible Lorentz asymmetry of the medium. In the comoving frame of the condensate, it is a good approximation to assume isotropy and parametrize
$\sigma^{\mu\nu}$ as\footnote{Any nonzero constant value for $\sigma^{00}$ can be absorbed into
the definition of $f_G^2$, so  $\sigma^{00}$ can be set to unity without
loss of generality.}
\begin{equation}
    \sigma^{00} = 1, \hskip .3in
    \sigma^{ij} = - \delta^{ij} (1- \kappa )\,,
    \label{syn9cd}
\end{equation}
where $\kappa$, like $f_G$ can be treated as a parameter of the low energy EFT.

The action in Eq.~(\ref{syn9c}) can be contrasted to the one  discussed in section~\ref{sec:CGC-EFT}, where the dynamics of  reggeized 
gluons in the CGC is obtained without including the color fluctuations in the condensate which is treated as being static. Eq.~(\ref{syn9c}) is the first nontrivial modification of the action
in Eq.~(\ref{CGC3}). We have derived it here in the comoving frame of the condensate and will discuss further its transformation to the lab frame in 
Sec.~\ref{sec:Goldstone-final}.

As we mentioned previously, the time scale $f_G$ is parametrically much longer than the time-scale for processes involving reggeized gluon modes with $k^+ < \Lambda^+$. This time scale was estimated in \cite{Dvali:2021ooc} by taking into account the breaking of translational  invariance by the condensate. If $\Phi$ is the classical gauge field corresponding to $\rho_{\rm diag}$, this breaking is represented by 
\begin{equation}
    \delta \Phi = \epsilon_{X} R_S \partial_{X} \Phi = \epsilon_{X} f_G \,,
\end{equation}
where $\epsilon_{X}$ is a constant representing the infinitesimal change in the field due to translations in the $X$-direction in the co-moving frame of the condensate.\footnote{Note that here $X$ is the collective coordinate 
representing the ``body-centered" average of the spatial dynamics of the degrees of freedom of the condensate. We will discuss in Sec.~\ref{sec:Goldstone-final} the transformation of Eq.~(\ref{syn9c}) from the body-centered comoving frame to the lab frame.}   Further, $R_S$ is the screening radius of the saturated glue configuration; it corresponds to the inverse of $\mu\sim Q_S$--see Eq.~(\ref{eq:Gaussian}). Thus the Goldstone dynamics is localized in a medium of size  
$R_S$.\footnote{As we will discuss in section~\ref{sec:Superfluid}, on time scales where Goldstone interactions become important, the quantum pressure exerted by the condensate will cause this localized lump to flow as a superfluid.} 

Thus on dimensional grounds one expects $\Phi= \sqrt{N}/R_S$, where $N=O(1/\alpha_S)$ is the occupancy of the classical field. We thus obtain the result 
\begin{equation}
    f_G = \frac{\sqrt{N}}{R_S}\equiv \sqrt{N}\,Q_S\,.
\label{f_Gestimate}
\end{equation}
Since $N\gg 1$ for $\alpha_S\ll 1$, this is a parametrically much larger scale than 
$Q_S$. As we will discuss, the decay of the condensate takes place through the scattering of Goldstones into collinear gluons. The time scale for this can be estimated to be 
\begin{equation}
    \tau_G = 1/n\sigma = Q_S^{-3} f_G^2  = \alpha_S^{-1} Q_S^{-1}\,,
\end{equation}
where we estimate the density of Goldstone modes $n=Q_S^3$ and the scattering cross-section $\sigma =1/f_G^2$. A very similar analysis was performed in \cite{Baier:2000sb}. The discussion there was however in the context of heavy-ion collisions; as a result of the longitudinal expansion of the overoccupied fields, one obtains instead $\tau_G\sim \alpha_S^{-3/2} Q_S^{-1}$.

The other parameter in the EFT discussion in $\kappa$, which characterizes the color screening scale of the condensate, which should be of the order of the saturation scale $Q_S$. We can therefore estimate  
\begin{equation}
    1-\kappa = Q_S^2/f_G^2 = 1/N \,,
\end{equation}
where the occupancy $N=O(1/\alpha_S)$. In the original formulation of the classical CGC EFT~\cite{McLerran:1993ka,McLerran:1993ni}, there was no distinction made between the color screening scale of each individual color configuration $\rho_{\rm diag}$ and the scale $\mu^2$ determining the statistical distribution of color configurations. However, as noted in \cite{Iancu:2001md} (see also \cite{Krasnitz:2002mn}), fundamentally, these are distinct scales, with the former representing the deconfined dynamics of the condensate. 

The estimates here are of course only parametric but can be quantified by systematic comparison of the EFT to experiment, just as in other low energy EFTs such as chiral perturbation theory. The takeaway from our discussion is that a consistent description of the condensate requires the Goldstone field $U$; its dynamics brings in a new scale $f_G$ in addition to the screening scale ($1-\kappa\sim Q_S$) in the effective field theory description of gluon saturation.

\section{Dynamics of Goldstone and collinear gauge fields}
\label{sec:Goldstone-dynamics}
The Goldstone action in Eq.~(\ref{syn9c}) describes the dynamics of $U$ and its coupling to the gauge field via the covariant derivative. But there is also the coupling of the gauge field to the charge density of the condensate. In the CGC EFT, this is described by the second term in Eq.~(\ref{CGC3}) which contains the logarithm of the path-ordered Wilson line. We will now discuss how this term can be rewritten in a simpler form using the parametrization of the charge density in terms of the condensate charges $\rho_{\rm diag}$ and the Goldstone fields $U$, as in Eq.~(\ref{syn9}). This reformulation will prove useful when we discuss the equations of motion of the novel Goldstone  and collinear gauge degrees of freedom.

Consider the Wilson line operator along a curve $C$,
\begin{equation}
W_{ij}(R, C) = \left[ \P \exp\left( \int d\tau\, (-i T_a) A^a_\mu \frac{d x^\mu}{d\tau}
\right)\right]_{ij}\,,
\label{syn4}
\end{equation}
where $T_a$ are matrices in the representation $R$.
The action in Eq.~(\ref{CGC3}) is written using $\ln W$, which is
rather involved because of the path-ordering in defining $W$.
While we can express $\ln W$  via a series expansion as in 
Eq.~(\ref{CGC5}), a simpler expression is possible by use of the additional
degrees of freedom in $U$. 
As mentioned in the Introduction, this can be done in terms of
the Kostant-Kirillov-Souriau (KKS) action which describes the transport of nonabelian charges along a worldline.
The goal is to write the Wilson line as a path integral with a
classical action where we have a set of functions $Q^a$ which
lead to $T_a$ upon quantization. As noted previously, this can also be achieved in the worldline framework employing Grassmanian and bosonic worldlines~\cite{Mueller:2019gjj} but we will find it more convenient here to work with the KKS action. The path-ordering is also simplified since one is dealing with $c$-number functions rather than the matrices $T_a$.
\subsection{KKS action and the Wilson line: General considerations}
\label{sec:KKS}
The basic KKS action is given by\footnote{The KKS action is central to the idea of geometric quantization and has been discussed in many books and reviews; some recent books include \cite{Woodhouse:1992de}, \cite{Sniatycki:1995sn}, \cite{Woit:2017vqo}, and \cite{Nair:2024wyq}.
}
\begin{equation}
S_{\rm KKS} = 2 i \int_{\tau_{\rm i}}^{\tau_{\rm f}}  d\tau~ \Tr \left[ w_s h_s \left( g^{-1} \partial_\tau g\right)
\right]\,.
\label{syn1}
\end{equation}
Here $g$ is an element of the compact Lie group $G$, taken as a matrix in the fundamental representation. We remind the reader that
$h_s$ are the diagonal generators of the Lie algebra of $G$
in the chosen basis.
In Eq.~(\ref{syn1}), $w_s$ are a set of real numbers.
Notice that the action is invariant under reparametrization of $\tau$.
This freedom can be used to set $\tau_{\rm i} = 0$ and
$\tau_{\rm f} = 1$ in Eq.~(\ref{syn1}). We also see that
the action changes only by boundary terms at $\tau_{\rm f}$ and
$\tau_{\rm i}$ under $g \rightarrow g u$ where $u $ is an element of
the Cartan subgroup ($u \sim e^{i h_s \theta_s}$). For this reason,
effectively, the theory is defined on the space $G/H$.

This action is relevant for us for the following reason. Its quantization
gives a finite dimensional Hilbert space of states corresponding to a unitary irreducible representation (UIR) of the group $G$. 
Therefore the KKS action can be used as
the classical description (and hence appropriate in a path integral)
of the color degrees of freedom for a quark or gluon or any collection thereof. The particular UIR which emerges from quantization of
Eq.~(\ref{syn1}) is chosen by the $w_s$. In fact, the set of numbers
$(w_1, w_2, \cdots, w_r)$, where $r$ denotes the rank of algebra, will be the highest weight of the representation. 
We will not discuss the derivation of this result here; 
some of the details of the quantization and the emergence of the
UIR are given in Appendix A.

Since the KKS action leads to a UIR of the group, it can be used
to construct the representation of the Lie algebra describing the colored states
and thence to obtain a path integral expression for 
the Wilson line.
Towards this end, we consider the action
\begin{equation}
S_{\rm KKS}(g, A) = 2  i \int d\tau~ \Tr  \bigl[ w_s h_s  g^{-1} \left( \partial_\mu g + (-it_a )A^a_\mu g\right)\bigr] \frac{dx^\mu}{d\tau}\,.
\label{syn4a}
\end{equation}
The $\tau$-derivative is now taken along the curve $C$ given by
$x^\mu\equiv x^\mu(\tau)$.
The first term in this action $S(g,A)$,
being of the first order in the $\tau$-derivative, does not contribute to the Hamiltonian. The second term involving the gauge field leads to a nonzero Hamiltonian. The corresponding
$\tau$-evolution operator can then be used to
construct  a path integral representation of
the Wilson line as
\begin{equation}
W_{ij}(R, C) = \int [d\mu(g)] ~ \psi_i(g,1)^*  \,\left[ e^{i S_{\rm KKS} (g, A)} \right]\, \psi_j(g,0)\,,
\label{syn5}
\end{equation}
where the action is defined as in Eq.~(\ref{syn4a}).
For more details, see Appendix A.

The action in the path integral in Eq.~(\ref{syn5}) is
a gauged version of the KKS action along the curve $C$.
We have taken $\tau$ to be defined in the interval $[0,1]$.
This action connects the initial value of $g$ to the final value of
$g$; the wave functions
$\psi_i (g,1)$ at $ \tau =1 $ and $\psi_j(g, 0)$ at $\tau=0$ then 
map the labeling of states by
$g$ to the colored states $\ket{i}$, $\ket{j}$.
Further, in Eq.~(\ref{syn5}), 
$d\mu (g)$ is the Haar measure for each value of $\tau$
for the coset space $G/H$. An important point is that one can extend this to the full $G$ if we divide out the volume of $H$ at each $\tau$. This is because $S$ changes by boundary terms under
$g \rightarrow g u$, but the corresponding phase factors are canceled out
by the phase changes in $\psi_i^*$, $\psi_j$.

Defining $j^\mu_s = w_s \frac{dx^\mu}{d\tau}$, the KKS action can be rewritten as
\begin{equation}
S_{\rm KKS} =  2 i \int d\tau~ j^\mu_s \Tr  \bigl[h_s  g^{-1} \left( \DU_\mu g \right)\bigr],
\hskip .3in \DU_\mu g = \partial_\mu g + (-it_a) A^a_\mu g\,.
\label{syn6}
\end{equation}
Here $j^\mu_s$ are currents for the diagonal charges $\rho_{\rm diag}$ along the worldline, with the action describing the transport of nonabelian charges along the worldline $C$.
Consider now a collection of worldlines $x^\mu (\tau_{(\alpha)})$
where $\alpha = 1, 2, \cdots, N$ label the worldlines.
The action then takes the form
\begin{equation}
S_{\rm KKS:sum} =  2 i \sum_\alpha \int d\tau_{(\alpha)}~ j^\mu_s \Tr  \bigl[h_s  g_{(\alpha)}^{-1} \left( \DU_\mu g_{(\alpha)} \right)\bigr]\,.
\label{syn6a}
\end{equation}
If we consider a collection of nearby worldlines, we can gather them as
a nearly continuous distribution. Then, as is the case in fluid dynamics,  we can
coarse-grain over this set of worldlines
using the position at a given $\tau$ to label them: 
$\sum_\alpha d\tau_{(\alpha)} \rightarrow d^4x$. 
This leads to the coarse-grained action 
\begin{equation}
S _{\rm KKS:CG}= 2  i \int d^4x~ j^\mu_s(x) \Tr  \bigl[h_s  g^{-1} \left( \DU_\mu g \right)\bigr]\,,
\label{syn7}
\end{equation}
where we have replaced the worldline current with a coarse-grained spacetime dependent current $j^\mu_s$. For a large number $N\gg 1$ of worldlines per phase space volume, this corresponds to the $O(1/g)$ color charge density discussed previously~\cite{McLerran:1993ka,McLerran:1993ni,Jeon:2004rk}. 
One can in principle add dynamics for the worldlines themselves.
For a single Wilson line, this would be like adding an action, say,
$m \int\sqrt{{\dot x}^2} d\tau$, in the Wong action.
More generally, we have to specify the action for $j^\mu_s$, as we will now discuss. 

\subsection{Relating KKS to the Goldstone fields}
\label{sec:KKS-Goldstone}
The use of the gauged and coarse-grained version of the KKS action
in Eq.~(\ref{syn7}) may seem like something of a mathematical excursion but its
appropriateness for our problem becomes clear from relating it
to the Goldstone modes. Consider the parametrization of the color 
charge density in terms of its diagonal form and the Goldstone modes
in Eq.~(\ref{syn9}). Taking the trace of that expression with $t^a$, 
the color components of the charge density are 
\begin{equation}
\rho^a = 2\Tr (t^a \rho) = 2\, \Tr (t^a U \rho_{\rm diag} U^{-1})\,.
\label{syn7a}
\end{equation}

At the same time,  taking
the derivative of the KKS action with respect to $A^a_+$, we identify the color charge density associated with the action as\footnote{Here, as in the CGC EFT, we are working in the infinite momentum frame, where ``+" denotes the large component of the hadron's momentum.}
\begin{equation}
{\delta S_{\rm KKS:CG} \over \delta A^a_+} = 2\, \Tr (t^a g j^+ g^{-1})\,,
\hskip .3in j^+ = j^+_s h_s\,.
\label{syn7b}
\end{equation}
Note that $j^+$ is diagonal. 
By comparing these two expressions, we see that
we can identify 
\begin{equation}
\label{eq:Goldstone+current}
    g = U\,\,\,\, {\rm and}\,\,\,\, j^+_s h_s = \rho_{\rm diag}\,.
\end{equation}
We will make this identification from now on. We observe further that $g$ in Eq.(\ref{syn7}) also belongs to $G/H$, just as it is for $U$--see the discussion in the paragraph after Eq.~(\ref{syn5}). Gauge transformations act as $g \rightarrow S g$, and Eq.~(\ref{syn7}) has the corresponding gauge-covariant derivative in it, with $A \rightarrow S A S^{-1} - \partial S \, S^{-1}$.
This is in agreement with what we found for $U$ in section \ref{sec:Goldstone}.

These observations amount to the statement that the coarse-grained KKS action in 
Eq.~(\ref{syn7}) with $g \rightarrow U$ represents the term with the log of the Wilson line in Eq.~(\ref{CGC3}); moreover, $U$ is indeed the Goldstone field discussed.
We conclude therefore that the last term in the action in Eq.~(\ref{CGC3}) in section~\ref{sec:CGC-EFT} can be replaced by
\begin{equation}
S_{\rm int.} = 2  i \int d^4 X~ j^\mu_s(X) \Tr  \bigl[h_s  U^{-1} \left( \DU_\mu U \right)\bigr]\,.
\label{syn7c}
\end{equation}
This term incorporates the coupling of the condensate to $A_\mu$ and to the Goldstone modes. 

There is naturally a path integral over $U$ which will involve
the action $S_{\rm Goldstone}$ in Eq.~(\ref{syn9c}) and this term $S_{\rm int.}$.
If we do not include
the kinetic term for $U$ from Eq.~(\ref{syn9c}),  the path integral over
$U$ reduces to the form given in Eq.~(\ref{syn5}), corresponding to a
collection of noninteracting Wilson lines. The summation over these
Wilson lines may be represented by integration with a density
$\rho^a$ and we recover the form of the last term
in Eq.~(\ref{CGC3}) with $\ln W$ coupled to $\rho$. This case without dynamics for $U$
was the scenario considered in the CGC EFT description of  \cite{Jalilian-Marian:2000pwi}. The KKS action was not needed there since one could integrate out
$U$ and deal directly with the Wilson lines.
When we have the Goldstone fields with a kinetic term $S_{\rm Goldstone}$ as
in Eq.~(\ref{syn9c}), this can no longer be done. We need the generalization given here.

Another feature of Eq.~(\ref{syn7c}) worthy of comment follows from the observation that  the color current is given by\footnote{We will soon see that there is another contribution to the current from the Goldstone action discussed below in section~\ref{sec:Goldstone-decay}.}
\begin{equation}
{\delta S_{\rm int.} \over \delta A^a_\mu}
=  2 j^\mu_s \Tr( h_s U^{-1} t^a U) = j^\mu_s Q^a_s\,,
\label{syn7d}
\end{equation}
for $Q^a_s = 2 \Tr (h_s U^{-1} t^a U)$, where we have the current due to the condensate alone. This part of the current has a factorized form
(for each value of $s$, where $s$ denotes the diagonal generators)
known as the Eckart form in the context of fluid dynamics.
To clarify its nature, note that $Q^a$ can be thought of as the color charge in the rest frame of a particle in the rest frame, whose Lorentz-transform has the form $Q^a {\dot x}^\mu$. For a collection of independent worldlines we would then have a similar form for the current density, with each charge drawn out into a worldline, which results in the expression above.  However, if there are interactions among the worldlines, the result need not have the Eckart form. Thus the form of $S_{\rm int.}$ in Eq.~(\ref{syn7c}) is in the approximation that the interactions among the constituent particles of the condensate are small compared to the
current generated by the overall flow of the condensate. As we have discussed, these interactions are quantified by the dynamics of the $U$ fields. 

\subsection{Goldstone decay into collinear gluons}
\label{sec:Goldstone-decay}
While the expression given in Eq~(\ref{syn7c}) is a possible term in the
effective action, it needs modification 
to include the absorption and emission of gluons by the condensate.
The reason for this can be seen from the following argument. Consider the equations of motion we obtain from Eq.~(\ref{syn7c}) by varying $\delta U = i U h_s \delta \theta_s$,
which is the infinitesimal version of $U \rightarrow U \, u$. The classical equations of motion then read 
\begin{equation}
\partial_\mu j^\mu_s = 0\,,
\label{syn10}
\end{equation}
which shows that the action as written leads to
the conservation of the diagonal charge current. In the CGC EFT, this current had only a ``+" (light front charge) component in the large $P^+$ direction; the dynamics of the Goldstone fields however ensures that there are nonzero components to this charge current.\footnote{These non-eikonal contributions to the current are typically thought of as being suppressed by $1/P^+$; our analysis shows that these are suppressed by the occupancy instead, which is $O(1/N)$.} Further, if the given charge distribution emits a gluon, there should be a change in the leftover color charge of the charge distribution. This means that a piece of $j^\mu_s$ must correspond to the gluon charge current while the rest remains as the color charge density of the condensate.

To identify the gluon part of the color
current which goes into the KKS action, we start by noting that
the Yang-Mills equations of motion
\begin{equation}
\partial_\mu F^{\mu\nu} = [F^{\mu\nu}, A_\mu] \,,
\label{gc1}
\end{equation}
show that there is a conserved current $[F^{\mu\nu}, A_\mu]$ obeying $\partial_\nu ([F^{\mu\nu}, A_\mu] ) = 0$. One may try to identify this with the color current of the gluons but, as is well-known, this is problematic since $[F^{\mu\nu}, A_\mu] $ is not gauge-invariant. It is in fact not even gauge-covariant. However, unlike the general situation, in the present case we have the Goldstone field $U$.
Notice that $U^{-1} F^{\mu\nu} U$ is gauge-invariant because, as we discussed previously, $U$ transforms as
$U \rightarrow S\,U$ and  $F^{\mu\nu} \rightarrow S F^{\mu\nu} S^{-1}$.
The Yang-Mills equations in Eq.~(\ref{gc1}) can then be reexpressed as
\begin{equation}
\partial_\mu ( U^{-1} F^{\mu\nu} U)  = [U^{-1} F^{\mu\nu} U , U^{-1} \DU_\mu U] \,,
\label{gc2}
\end{equation}
with r.h.s now defining a current which is
both conserved and gauge-invariant:
\begin{equation}
(-i t_a) J^{a \nu} = J^\nu = [U^{-1} F^{\mu\nu} U , U^{-1} \DU_\mu U] 
= U^{-1} \bigl( [F^{\mu\nu} , \DU_\mu U U^{-1} ] \bigr) U\,.
\label{gc3}
\end{equation}
The field $U$ was defined by the diagonalization of
the charge density $\rho$. We can therefore view it as defining
the local frames for the color orientation in which 
$\rho$ has only the diagonal components; recall $\rho_{\rm diag} = U^{-1}\rho\, U$.
Along the same lines, with the conjugation by $U$
in Eq.~(\ref{gc3}), we have the
projection of a gauge-covariant density $[F^{\mu\nu} , \DU_\mu U U^{-1} ]$
to the same set of color frames.
Now consider adding the charge density $J^+$ to the 
condensate.\footnote{This is also the case in the CGC RG framework when the scale separating the color charge density from the reggeon field is changed from $\Lambda^+\rightarrow {\Lambda^\prime}^+ $, thereby incorporating a piece of the gauge field in the condensate at the new scale ${\Lambda^\prime}^+$ .}  The new density will be 
\begin{equation}
\rho_{\rm diag} \rightarrow \rho_{\rm diag} + J^+
= \rho_{\rm diag} + J^+_s h_s + J^+_A E_A
= \rho'_{\rm diag} + J^+_A E_A\,,
\label{gc4}
\end{equation}
where we have expanded $J^+$ in terms of the generators of the group.
Since the condensate carries a large charge, 
the absorption and emission of gluons will not change it significantly,
so $\rho'_{\rm diag}$ will still be the dominant term in
Eq.~(\ref{gc4}). This will be a good approximation until the condensate is
significantly depleted.

Likewise, the current corresponding to $\rho_{\rm diag}$ is modified by 
\begin{equation}
j^\mu_s \rightarrow j^\mu_s + J^\mu_s
= j^\mu_s + 2 i \Tr
\left( h_s [U^{-1} F^{\mu\nu} U , U^{-1} \DU_\mu U] \right)\,,
\label{gc5}
\end{equation}
and Eq.~(\ref{syn7c}) is generalized to 
\begin{equation}
S_{\rm int:gen} = 2 i  \int d^4X\,
\Bigl( j^\mu_s + J^\mu_s
\Bigr) \, \Tr \bigl(h_s U^{-1} \DU_\mu U\bigr)\,.
\label{gc6}
\end{equation}
With the addition of $J^\mu_s$, 
the conservation law for the color current will be modified.
Using the variation $\delta U = i U h_s \delta \theta_s$,
we can see that the equation of motion for the color current now takes the
form
\begin{equation}
\partial_\mu (j^\mu_s + J^\mu_s ) = 0\,.
\label{gc7}
\end{equation}

To clarify the construction of the EFT better, we start by recalling
that there are two kinematic regimes of interest for the gluons.
The local perturbations in the color orientation of the condensate
are described by the Goldstone bosons. The condensate can also emit gluons which are nearly collinear with the condensate with their $k^+$-values in a small range $\Lambda^+ < k^+ \leq P^+$.
We will use $\A_\mu$ to denote these collinear
gluons. The condensate also creates the classical field of the shock wave discussed in section~\ref{sec:CGC-EFT}, responsible for the interaction of the probe with the condensate on short time scales $x^+\sim 1/Q$.
This field leads to the ladder of reggeized gluons described by the BFKL RG equation and its nonlinear generalizations. We will denote it by $A_\mu$.
The distinction between $A_\mu$ and $\A_\mu$ is one of time scales,
with $\A_\mu$ being relevant on the long time scale of the decay of the
condensate. 

The Yang-Mills action will in principle have terms which mix $A_\mu$ and
$\A_\mu$ reflecting the interactions between the two types of gluons.
In the effective action we are constructing, we neglect these as subleading on the time scales of the absorption and emission of collinear gluons by the condensate. Since this occurs on a longer time scale compared to the dynamics of the reggeized gluons $A_\mu$, the significant depletion of the condensate (whose color charge density is of 
order $\sim Q_S^3$) will take a much longer time compared to the time scale for the reggeized gluon emissions. It is even large compared to $\tau_G$, because
of the large value of $Q_S^3$ in Regge asymptotics.
It therefore follows that due to the dynamical separation of time scales, the gauge potentials within the covariant derivative of the current $J^\mu_s = 2 i \Tr
\left( h_s [U^{-1} F^{\mu\nu} U , U^{-1} \DU_\mu U] \right)$
should refer to the collinear gluons $\A_\mu$ only.

\subsection{Putting it together: EFT describing color Goldstones, collinear glue and condensate decay}
\label{sec:Goldstone-final}
We will now collect various terms in the effective action 
that we have discussed. Before we do that, there is one more modification we will need.
The KKS action, as well as the Yang-Mills equations of motion were discussed 
in the lab frame, while the action for the Goldstone modes was expressed in the comoving frame of the condensate. As shown in Appendix B, the latter can similarly be expressed in the lab frame, first by straightforward Lorentz transformation and then reexpressing the
flow velocity in terms of the currents $j^\mu_s$.
The result derived there is 
\begin{eqnarray}
S_{\rm Goldstone} 
&=&-{1\over 2} f_G^2\int d^4 X\,\,
{\tilde \sigma}^{\mu\nu} 
\sum_{\talpha} \Tr \left( E_{\bf\talpha}
U^{-1} \DU_\mu U\right) \Tr \left( E_{-\bf\talpha} U^{-1} \DU_\nu U\right)\,,
\nonumber\\
    {\rm with}\,\,\,{\tilde \sigma}^{\mu\nu} &=&
    \sum_s {1\over j^2_s} \left( \eta^{\mu\nu} (1- \kappa) j^2_s + \kappa j^\mu_s j^\nu_s
\right)\,.
\label{disc0}
\end{eqnarray}

This action for the Goldstone modes is the principal novel ingredient in generalizing the CGC action in 
Eq.~(\ref{CGC3}). We can expand this expression in terms of unconstrained Goldstone fields similarly to what is done in the chiral Lagrangian for pseudoscalar mesons. 
Since the action is invariant under 
$U \rightarrow U u$, where $u$ is in the Cartan subgroup, we have six Goldstone fields
for the $SU(3)$ case, corresponding to the coset group $SU(3)/[U(1) \times U(1)]$.
As discussed earlier in section \ref{sec:Goldstone}, the generators of
$SU(3)$ can be split into the coset basis given by
$E_A$, $A = 1, 2, 4, 5, 6, 7$ and $h_s$ for the diagonal
generators $t_3$, $t_8$.
We can choose a parametrization for $U$, 
\begin{equation}
U = \exp\left( i  {2 \over f_G} \varphi^A E_A \right)\,,
\label{disc1}
\end{equation}
where $ \varphi^A$ represent the Goldstone fields and the prefactor
$2/f_G$ in the exponent is chosen so that the leading term 
in the action has the properly normalized form  ${1\over 2} (\partial \varphi)^2$.
Working out the expansion to the quartic order in the fields is straightforward. 
(Some of the details are given in Appendix B.) The result is
\begin{eqnarray}
\label{eq:colored-pions}
    S_{\rm Goldstone} &\rightarrow &  {1\over 2} \int d^4 X\,\,{\tilde \sigma}^{\mu\nu}\biggl[
\left(\partial_\mu {\varphi}^A - {f_G \over 2} \A^A_\mu\right)
\left(\partial_\nu {\varphi}^A - {f_G \over 2} \A^A_\nu\right) \nonumber \\
&+& 2 f^{sBC} \A^s_\mu {\varphi}^B \partial_\nu {\varphi}^C + \A^A_\mu f^{A BC} \varphi^B \partial_\nu \varphi^C
- f_G f^{sBC} \A^s_\mu \varphi^B \A^C_\nu \nonumber \\
&+& {1\over f_G^2} 
f^{A BC} f^{A DE} {\varphi}^B \partial_\mu {\varphi}^C \, {\varphi}^D \partial_\nu {\varphi}^E 
- {4\over 3 f_G^2} f^{l BC} f^{l DA} \varphi^B \partial_\mu \varphi^C \varphi^D 
\left(\partial_\nu \varphi^A - {f_G \over 2} \A^A_\nu\right)\nonumber \\
&-& f^{s BC}  f^{s DE} \varphi^B \A^C_\mu \varphi^D \A^E_\nu -f^{sBC} f^{AaC} \A^s_\mu \varphi^A \varphi^B \A^a_\nu
+ {2 \over f_G} f^{sBC} f^{sDE} \varphi^B \partial_\mu\varphi^C \varphi^D \A^E_\nu + \cdots \biggr]\,.
\end{eqnarray}
There are several interesting features of this (up to $O(\varphi^4)$) expansion of the Goldstone action. Firstly, the minimal coupling of the colored Goldstone fields to the collinear gluons contains a mass term for the latter, which must be interpreted as the in-medium screening mass. Indeed, this mass is exactly analogous to the photon mass (or inverse London penetration depth) in the Anderson-Higgs low-energy effective action for the broken symmetry phase of a superconductor and similarly, has a rich structure.  

Further, we grouped the interaction terms in Eq.~(\ref{eq:colored-pions}) in terms of powers of 
$p/f_G$. The terms in the second line are all of order $O(p/f_G)$ and include the three-vertex for a colored Goldstone to absorb or emit a collinear gluon, as well as that for two collinear gluons to emit or absorb a colored Goldstone. The terms on the third and fourth lines are of $O((p/f_G)^2)$; these suppressed interactions correspond to four-particle vertices involving the self-coupling of the Goldstone modes, contact interactions of two Goldstones and two collinear gluons, as well as the coupling of three Goldstones to a collinear gluon. Note further that the interaction terms for collinear gluons emitted in the condensate color directions are distinct for those in the color directions corresponding to the Goldstones.

We can now collect the terms 
in the effective action which include, besides the Yang-Mills action, the contributions from Eqs.~(\ref{syn9c}) and (\ref{gc6}):
\begin{eqnarray}
S_{\rm CGC+fluid} &=& {1\over 2} \int d^4 X\, \Tr 
(F_{\mu\nu})^2  
-{1\over 2} f_G^2\int d^4 X\,\,
{\tilde \sigma}^{\mu\nu} 
\sum_{\talpha} \Tr \left( E_{\bf\talpha}
U^{-1} \DU_\mu U\right) \Tr \left( E_{-\bf\talpha} U^{-1} \DU_\nu U\right) \nonumber\\
&+&  2 i  \int d^4 X\,
\Bigl( j^\mu_s + J^\mu_s \bigr)
\Bigr) \, \Tr \bigl(h_s U^{-1} \DU_\mu U\bigr)\,,
\label{syn18}
\end{eqnarray}
where $J^\mu_s$ is given in Eq(\ref{gc5}), with the $A$ fields in 
$J^\mu_s$ restricted to the collinear gluons $\A_\mu$.

This action has all the right symmetries. 
 The group parameters in $U$ referring to the coset $G/H$
 will play the role of the ``Goldstone"
 particles corresponding to low energy excitations of the condensate.
 The second term in $S$ proportional to $f_G^2$ is the usual kinetic  term for
 such excitations.
The term $ j^\mu_s \Tr [h_s U^{-1} \DU_\mu U] $ expresses the Wilson line
in the form of an action {\it \`a la} KKS. It is the generalization of 
$\Tr \rho \ln W$ as in Eq.~(\ref{CGC3}) introduced previously in \cite{Jalilian-Marian:2000pwi}.
We also have the additional term 
$2i J^\mu_s\, \Tr [h_s U^{-1} \DU_\mu U] $ which accounts for the absorption/emission of collinear gluons by the condensate.
 
Finally, since $\rho_{\rm diag}$ is not a dynamical degree of freedom in the action, it can be treated as the static source of color charge that the dynamical degrees of freedom in the above action ($A, U, \A$) couple to. However, in Eq.~(\ref{eq:CGC-path-integral}) one must make the modification
\begin{equation}
    \int [d\rho] F[\rho]\rightarrow \int [d\rho_{\rm diag}]\, F[\rho_{\rm diag}]\,,
\end{equation}
to represent the stochastic probability distribution over classical color charge densities which is well approximated by the Gaussian weight functional with the variance $\mu^2\sim Q_S^2$. This states simply that on average an external probe will scatter off color neutral configurations but will also be sensitive to configurations with large color charge. Note that the elimination of $\rho_{\rm diag}$ leads to a term of the form $\bigl( \Tr [h_s U^{-1} \DU_\mu U]\bigr)^2$. This term serves as the kinetic term for the Cartan subgroup part of $U$. Such a term is also what we expect for fluid dynamics, as we will discuss next in section~\ref{sec:Superfluid}.

Though our focus has been on kinetic terms in the Goldstone action, it can can include higher derivative contributions in the $U$ fields with corresponding ``low energy constants", as well as topological terms, just as in the case of chiral perturbation theory. The dynamics of the color glass condensate therefore in principle includes further structure that can be uncovered by careful studies of the fragmentation region of the hadron/nucleus in high energy experiments. 
\section{From Color Glass to Color Superfluid }
\label{sec:Superfluid}
The CGC-fluid action in Eq.~(\ref{syn18}) has a natural superfluid interpretation. 
Recall that superfluidity can be viewed as being due to
the condensation of particles into the ground state leading to
the spontaneous breaking of the particle number symmetry
(as in liquid Helium) or isospin symmetry (as for pion superfluids).
This can be compared to the formation of the CGC and the
breaking of $G$ to the Cartan subgroup $H$. 
The Goldstone modes in $G/H$
can be viewed as forming the collective superfluid excitations, with
the condensate specified by the value of $\rho_{\rm diag}$.
This is analogous to pion condensation and the breaking of isospin 
symmetry, except that the relevant group is the color group in 
our context.

The dynamics of the Goldstone modes in $G/H$ will also lead to changes
in the current corresponding to the Cartan subalgebra. 
This is clear from the fact that one has only one 
combined conservation law, that given in Eq.~(\ref{gc7}).
Thus there is another superfluid component corresponding to
the diagonal directions whose dynamics is governed by 
the KKS action and for which superfluidity in
liquid Helium provides a closer analogy. 

Consider 
the action for a bosonic Schr\"odinger field
given by
\begin{equation}
    S = \int \left[ i \psi^\dagger {\partial \psi \over \partial t}
    - \int {{\nabla\psi^*} {\nabla\psi}\over 2M}  - \int \psi^*\psi(x) V(x,y) 
    \psi^* \psi(y)\right]\,,
    \label{syn31}
\end{equation}
which can be applied to liquid Helium, with $M$ specifying the mass of the Helium atom. 
Using the Madelung parametrization $\psi = \sqrt{\rho}\, e^{-i \theta}$, 
this action reduces to the form
\begin{eqnarray}
    S &=& \int \left[ {\rho} \, {\dot \theta} - {\rho \over 2 M }(\nabla \theta )^2
      - f (\rho ) \right]\nonumber\\
      &=&\int \left[  \rho \, {\dot \theta} + {\vec j} \cdot \nabla \theta
      + { M {\vec j}\cdot {\vec j} \over 2 \rho} - f(\rho) \right]
\label{syn32}
\end{eqnarray}
The second  form of this action leads back to the first form upon elimination of ${\vec j}$
by its equation of motion.
Notice that the leading term in this second form (Eq.~(\ref{syn32})) of the action 
is $i j^\mu g^{-1} {\partial_\mu g}$
where $g = e^{-i \theta}$ is an element of $U(1)$ and $j^\mu = (\rho, {\vec j})$. Therefore it indeed has the form of the coarse-grained KKS action in
Eq.~(\ref{syn18}). The situation is
simpler in the context of Helium since we are dealing with a $U(1)$ symmetry group. We note parenthetically that since this example is in a nonrelativistic context,
a comparison of the second set of terms in
Eq.~(\ref{syn32}) with some of the other terms in Eq.~(\ref{syn18}) is not relevant.

The action in Eq.~(\ref{syn18}) can also be viewed in
the light of a general group-theoretic approach to fluids.
To bring out the connection, consider a fluid comprised of a large number of
point particles, each of which is charged under a symmetry group $G$.
Thus the possible states of the particle will correspond to unitary irreducible representations of $G$.
Since it is the KKS action, which upon quantization gives a UIR of a Lie group
$G$, the transport of the charged particles is described by
such an action. For a collection of particles, one can then consider a coarse-grained version of the KKS action. 

Thus fluid dynamics of any kind, based on a particulate substratum, should admit a group-theoretic formulation using the KKS action.
For instance, ordinary relativistic fluid dynamics
would correspond to a coarse-grained KKS action for the Poincar\'e group
\cite{Karabali:2014vla}. This can also describe mass and spin transport since they are the charges for the Poincar\'e group.
For colored particles with $G = SU(3)$, the corresponding fluid dynamics was considered
previously in \cite{Bistrovic:2002jx},\cite{Jackiw:2004nm}, \cite{Nair:2011mk}, leading to the action
\begin{equation}
S = 2 i \int d^4x \, j^\mu_s \, \Tr \left( h_s U^{-1} D_\mu U\right)
- \int d^4x\,F (j^\mu_s j_{\mu s'}) \,,
\label{syn28}
\end{equation}
where $U$ is now an element of the color group.
The first term in this action is identical to the similar KKS term in 
Eq.~(\ref{syn18}).

We should point out that not all the terms in
Eq.~(\ref{syn18}) are obtained in the analogous examples mentioned above.
For example, the action in Eq.~(\ref{syn18}) has the term
\begin{equation}
2 i \int d^4x\, J^\mu_s \Tr (h_s U^{-1} \DU_\mu U )
= 
2i \int d^4x\, \Tr \bigl(h_s [ (U^{-1} D_\nu U ), (U^{-1} F^{\nu\mu} U)]\bigr) \, \Tr \bigl(h_s U^{-1} D_\mu U\bigr).
\label{syn29}
\end{equation}
This is not present in (\ref{syn28}), but it
is crucial for describing the possible decay of
the condensate by the emission of gluons.
We also note the difference in how the term 
$F (j^\mu_s j_{\mu s'}) $ is used. In Eq.~(\ref{syn28}), this is another term
in the action which leads to equations for $j^\mu_s$.
In contrast, in the CGC context, we integrate over all fields in 
Eq.~(\ref{syn18}) to calculate correlators as functions of $j^\mu_s$
and then a stochastic averaging over $\rho_{\rm diag}$ is carried out at the final step.

Moreover, the group theoretic description of colored nonabelian fluids did not address the case of the breaking of the global symmetry discussed here, which leads to the superfluid interpretation of Eq.~(\ref{syn18}). Uncovering the detailed structure of this novel CGC superfluid is a task we will address in follow-up work.

\section{Summary and Outlook}
\label{sec:QV-CGC}
In the Regge asymptotics of $2\rightarrow 2+n$ scattering, BFKL evolution drives the system to a stochastic distribution of gauge-invariant colored states of maximal overoccupancy $N$ of $O(1/\alpha_S(Q_S))$, with color screening radius $Q_S(x) \gg \Lambda_{\rm QCD}$ and $\alpha_S(Q_S)\ll 1$. These lumpy nonperturbative semi-classical states of saturated glue are described by the CGC effective field theory. This EFT has developed into a powerful framework that has been applied to systematic computations of scattering processes ranging from deeply inelastic collisions to heavy-ion collisions at collider energies. In this work, we revisited the construction of the CGC EFT by carefully examining the global symmetries that are broken in the creation of such metastable semi-classical colored states. 

Following a general discussion of the interpretation of color charged states as Noether charges of a global $SU(N_c)$ symmetry group (denoted by $G$), we showed that the construction of gauge invariant states with large color charge requires the spontaneous breaking (in the large occupancy limit $N\rightarrow \infty$) of this symmetry, resulting in the formation of the color charged condensate. For $N_c = 3$, the  color charge density $\rho_{\rm diag}$ of the condensate has a residual global $U(1)\times U(1)$ symmetry of the Cartan subalgebra (denoted by $H$) while the colored Goldstone modes corresponding to the broken generators (satisfying the coset G/H algebra) are represented by the fields $U$. These fields transforming under $G$ satisfy the left symmetry $U\rightarrow gU$ in their coupling to gauge fields that transform as $A\rightarrow g Ag^{-1} -\partial g g^{-1}$. 
 
We next wrote down the kinetic term for the corresponding Goldstone action. Analogously to chiral perturbation theory, this construction introduces a dimensionful Goldstone decay constant $f_G$ into the CGC EFT.  Importantly, while the shockwave in the CGC is treated as a static lightcone mode (in $x^+$), the Goldstone action introduces novel dynamics to the condensate controlled by the Goldstone scale, with the temporal scale controlling the onset of this dynamics given by $\tau_G \sim 1/(\alpha_S Q_S)$. 

In addition to the breaking of the global $SU(N_c)$ symmetry, the overoccupied coarse-grained CGC state breaks Poincar\'e invariance. This is represented by the tensor $\sigma^{\mu\nu}$ in the Goldstone effective action. In the comoving frame of the condensate, the spatial screening dynamics is represented by its rotationally invariant components $\sigma^{ii}=-\delta^{ij}(1-\kappa)$, where we identify $1-\kappa\sim Q_S$ as the color screening scale in the CGC EFT; likewise, the temporal component $\sigma^{00}=1$, with the normalization to unity set by $f_G$. As long as $\alpha_S\ll 1$, boost invariance is a good assumption at small $x$, allowing one to simply boost the Goldstone dynamics to the laboratory frame of the scattering. 

As in chiral perturbation theory, one can expand the Goldstone fields $U$ into (colored) ``pion" fields, whose interactions with each other, and with nonabelian guage fields are then manifest, and ordered in powers of the effective coupling $p/f_G\sim Q_S/f_G$, which is parametrically $O(1/\sqrt{N})$. 
We observed further that the minimal coupling in the kinetic term $\partial_\mu \varphi - \frac{f_G}{2} \A_\mu $ is indicative of dynamics analogous to that of the broken (Anderson-Higgs) phase in superconductivity. Similarly, the collinear gauge fields $\A_\mu$ of the condensate can be understood as acquiring mass by ``eating up" the massless colored Goldstones; this mass can likewise be interpreted as the inverse of the London penetration depth of the color charged CGC superfluid.  The corresponding mass gap is $O(1/\sqrt{N})$ so it takes very little energy for these collinear gluons to be emitted from the condensate and in principle interact with the ``normal fluid" reggeized gauge fields $A_\mu$ emitted at earlier times $\sim 1/Q_S$. 

While the Goldstone action captures the rich in-medium dynamics of color Goldstone excitations and the collinear gauge fields they couple to, we also have to consider their coupling to the condensate $\rho_{\rm diag}$. We showed that this coupling is well-described by a coarse-grained gauge-invariant KKS action, which is a generalization of the gauge-invariant coupling of the color charge density in the CGC. Indeed, on time scales $\tau\sim 1/Q_S \ll \tau_G$ where the temporal dynamics of the $U$ fields is suppressed, one recovers the usual ``Wong term" ${\rm Tr}(\rho \ln(W))$ term of the CGC that we discussed in section~\ref{sec:CGC-EFT}. 

The impact of the KKS generalization of the Wong coupling term in  CGC actions is as follows. Firstly, it induces a contribution to $\rho_{\rm diag}$ (or more generally, the current $j_s^\mu$) from the emission and absorption of collinear gluons, specified by the current $J_\mu^s$. It is important to note that this change is completely distinct in time scales from the change in the condensate induced by the quantum dynamics of the reggeized gluons described by the usual BK/JIMWLK RG evolution equations of the CGC. Importantly, we see explicitly that the sum of the two currents is conserved. Thus the depletion of collinear gluons through excitations of the condensate in the scattering process, will lead over a longer time scale to the depletion of the condensate itself. 

Our generalization of the CGC EFT points to a two-superfluid picture of the temporal dynamics of the colored condensate. As we noted, the dynamics of the Goldstone fields themselves can be understood as being analogous to the low energy dynamics of a charged superconductor. This dynamics occurs primarily in the flow of the color charges corresponding to the broken generators. Because the flow also occurs in the spatial directions, it in turn induces such flow in the color charges 
belonging to the Cartan subalgebra. This is specified by the KKS action which determines the interactions of the condensate with the Goldstone and collinear gauge fields. The superfluid behavior of these ``diagonal" color charges is akin to that of a neutral superfluid, following from a Madelung-type formulation of the KKS equations of motion. 

There is much to explore to fully understand the rich structure of this novel superfluid.\footnote{A useful analogy is that of the dynamics of charged superfluids in QCD matter at large baryon or isospin chemical potential~\cite{Alford:2007xm}. In the CGC, $Q_S$ is the analog of the chemical potential for an individual large color charge configuration. Of course, one has here  a stochastic distribution of such ``chemical potentials". } 
Of particular interest will be the interactions of the phonons/plasmons and collinear gluons constituting the excitations of the condensate with the normal fluid reggeized gluons that are emitted by the condensate on much shorter time scales ($\sim 1/Q_S$), whose dynamics is described in the usual CGC EFT framework. We will briefly address here possible interesting implications in the context of DIS and in hadron-hadron/heavy-ion collisions.  

Let us consider first the DIS case, where at high energies (small $x$) a virtual photon probe scatters off a hadron (or nucleus). As we discussed earlier, this interaction takes place through the virtual photon fluctuating into a quark-antiquark ``dipole", which then scatters off the static CGC shockwave, whose color fields are screened on a distance scale $\sim 1/Q_S$. The RG picture arises from considering how this picture changes if we boost the virtual photon further by a rapidity $Y\sim 1/\alpha_S$ making it energetically favorable to emit a gluon, which can then further interact with the shockwave.\footnote{We will working in the infinite momentum frame where the virtual photon with large momentum $q^-$ and the hadron with large momentum $P^+$ are boosted towards each other such that the squared center of mass energy is $s=2q^-P^+$.}.  

In an ``aligned jet" configuration where, say, the quark carries a large fraction $z\sim 1$ of the virtual photon energy, it will interact with color charges that are static on the time scale of the interaction. On the other hand, the ``slow" anti-quark with smaller energy fraction $z$ will be sensitive to the dynamical evolution of the color charges. The problem is no different conceptually from taking into account sub-eikonal $1/P^+$ effects in 
the scattering of the aligned jet~ see for instance \cite{Agostini:2025vvx,Altinoluk:2022jkk,Altinoluk:2025ivn}. The essential difference is that the dynamics of the color charges here a $1/N$ effect, and therefore, in principle, more important when considering corrections to the static shockwave picture. 

In \cite{Ball:2018prg}, it was argued that the color rotations of the quark and the antiquark in dipole scattering off the static CGC shockwave could be interpreted as a signature of a color memory effect~\cite{Pate:2017vwa} , analogous to the gravitational memory effect~\cite{Strominger:2014pwa,Strominger:2017zoo}. It was understood in this context that the formation of the classical shockwave corresponds to the breaking of an asymptotic global symmetry. The color memory effect was interpreted as the physical relative displacement of the quark-antiquark pair. In the high energy scattering context, however, such a displacement could also arise from simple multiple scattering. As we have argued here, the breaking of the asymptotic global $SU(N_c)$ symmetry leads to nontrivial Goldstone dynamics. Therefore, it is rather the imprint of this dynamics on final states through which one can uniquely 
extract evidence for color memory. 

Our more general formulation of the CGC EFT can also impact the nonlinear BK/JIMWLK evolution equations beyond leading logarithmic accuracy in $x$. To appreciate this, consider the emission of a ``slow" gluon with frequency of  order $\alpha_S Q_S$ from the dipole, with a scattering lifetime $\tau\sim 1/k^- = 1/(\alpha_S Q_S)$. This reggeized gluon will also be sensitive to the decay of the shockwave. A concrete manifestation of this will be in the modificaton of Eqs.~(\ref{eq:shockwave-propa})-(\ref{eq:shockwave-vertex}). This has been considered previously in \cite{Altinoluk:2021lvu} for the case of a slowly varying time-dependent reggeon gauge field. A further consequence, as we have noted, is that the condensate will radiate collinear gluons that can have ``final state interactions" with gluons emitted by the dipole. These contributions can also impact the RG evolution equations at high orders. 

A particularly intriguing potential application is to diffractive DIS final states that are characterized by a ``rapidity" gap in the spectrum. In this case, it is a color singlet combination of reggeized gluons that interacts with the condensate. As we have seen in 
Eq.~(\ref{eq:colored-pions}) for the interactions of two collinear gauge fields, the mass gap for their emission into the rapidity gap may be larger. One may thus speculate that the ratios of rapidity gaps could potentially provide insight into the mass gap of the condensate.  

We now turn to consider high multiplicity final states in hadron-hadron and heavy-ion collisions. In the CGC framework, this is formulated as the collision of two semi-classical gluon shockwaves, where Eq.~(\ref{eq:YM-shockwave}) is sourced by two lightcone currents, 
with $J^\nu = \delta^{\nu+}\delta(x^-)\rho_A(x_\perp)+\delta^{\nu-}\delta(x^+)\rho_B(x_\perp)$. 
In the past light cone of the collision, the gauge fields are pure gauges, as specified by Eqs.~(\ref{eq:YM-Lorentz})-(\ref{eq:classical}). The sum of two pure gauges is not a pure gauge however, leading to large gluon field strengths in the forward light cone~\cite{Kovner:1995ja,Kovner:2001bh,Kovchegov:1997ke}; while a complete analytical solution is not known, the Yang-Mills equations have been solved numerically both for the 
boost-invariant 2+1-D case~\cite{Krasnitz:1998ns,Krasnitz:2000gz,Krasnitz:2002mn} and the full 3+1-D case~\cite{Berges:2013eia,Berges:2013fga,Berges:2014yta}. The overoccupied matter created in such collisions is often referred to as the Glasma; for a detailed review of the Glasma and its evolution into the quark-gluon plasma, see \cite{Berges:2020fwq}. 

The dynamics described in the Glasma framework is that of a normal fluid, albeit far-off-equilibrium. How is this modified in our generalization? In the collision of two classical lumps with energy $N\,Q_S$, the ratio of the gap size relative to the energy that separates the Goldstone excitations from that the condensate is $O(1/\sqrt{N})$; this leads to these collective excitations being released very rapidly and becoming part of the normal fluid. Because the condensate state $\rho_{\rm diag}$ is a genuine infrared degree of freedom, it does not influence the bulk dynamics of the normal fluid. The latter is well-described in the bottom-up thermalization scenario~\cite{Baier:2000sb}. 

There are two interesting possible variants to this scenario. One is that the dynamical evolution of the overoccupied normal fluid can source the condensate. Real-time simulations of scalar field theories with ``Glasma-like" overoccupied initial conditions show that the system develops an infrared cascade that leads to superfluidity~\cite{Deng:2018xsk}; discussiona in the Yang-Mills context can be found in \cite{Berges:2023sbs,Stasto:2026xqw}. The other concerns small systems or asymmetric (such as photon-nucleus or proton-nucleus) collisions, where the energy transferred from the probe may not be sufficient to overcome the mass gap. In that case, the flow observed could be the flow of the superfluid condensate. This last possibility is intriguing given the so-called ``ridge-like" flow pattern observed in such small systems at collider energies~\cite{Dusling:2015gta,Grosse-Oetringhaus:2024bwr}. 

Given the copious and precise collider data that is available, and anticipated in future at the Electron-Ion Collider, a careful comparison of the generalized CGC EFT we have developed here can help fix the parameters of the EFT, providing insight both into the many-body dynamics of gluon saturation and towards a deeper understanding of the nature of 
collective flow at collider energies.


\section*{Acknowledgements}
RV would like to thank Juergen Berges, Guillaume Beuf, Paul Caucal, Edmond Iancu, Larry McLerran, Yacine Mehtar-Tani, Farid Salazar, Bjoern Schenke, Anna Stasto, Iain Stewart and Varun Vaidya for discussions on subtle aspects of the CGC EFT. He would in particular like to thank Al Mueller for sharing his notes and for reinvigorating his interest in the topic. We would also like to thank Sam Bateman, Gia Dvali, Victor Gorbenko, Monica Pate, Anna Raclariu, Himanshu Raj, Ira Rothstein, Andy Strominger, Derek Teaney and Neil Turok for discussions on the broader context of this work. RV thanks Wei Li and Juergen Schukraft for discussions on puzzles in multi-particle production at collider energies.

VPN thanks the Physics Department of the Brookhaven National Laboratory for hospitality during several visits over the course of this work. VPN's research was supported in part by the U.S.\ National Science Foundation grant PHY-2412479.

RV is supported at BNL  by the U.S. Department of Energy, Office of Science under contract DE-SC0012704 and within the framework of the Saturated Glue (SURGE) Topical Collaboration in Nuclear Theory. RV is supported at Stony Brook by the Simons Foundation as a co-PI under Award number 994318 (Simons Collaboration on Confinement and QCD Strings). RV thanks the UK Royal Society and the Wolfson Foundation for a Visiting Fellowship and the Higgs Center at the Univ. of Edinburgh for their kind hospitality.
\section*{Appendix A: Quantizing the KKS action}
\def\theequation{A\arabic{equation}}
\setcounter{equation}{0}
In this appendix, we will briefly review how the quantization
of the KKS action leads to the space of colored states and 
the path integral for the Wilson line.
The KKS action in Eq.~(\ref{syn1}) is 
\begin{equation}
S_{\rm KKS} = 2 i \int_{\tau_{\rm i}}^{\tau_{\rm f}}  d\tau~ \Tr \left[ w_s h_s \left( g^{-1} \partial_\tau g\right)
\right]\,.
\label{A:syn1}
\end{equation}
As mentioned there, the Hilbert space obtained by quantizing this theory
is finite dimensional and corresponds to 
a unitary irreducible representation (UIR) of the group $G$, 
which can be identified as the color charge states of a system.
In other words, KKS is the classical action for the color algebra in the quantum theory.

Under $g \rightarrow g u$, where $u = e^{i h_s \theta_s} $ is in
the Cartan subgroup $H$, the KKS action changes as\footnote{We will omit the subscript KKS in the rest of this appendix.}
\begin{equation}
S(g u) = S(g) -   w_s \int d\tau\, {\dot \theta_s}
= S(g) - w_s \theta_s \Big\vert_{\tau_{\rm f}} + w_s \theta_s \Big\vert_{\tau_{\rm i}}\,.
\label{A:syn2}
\end{equation}
Since the Lagrangian changes by a total derivative, the latter integrates to
terms defined at the end points; the equations of motion are therefore unaltered by
the transformation $g \rightarrow g u$.
Since the classical dynamics is insensitive to this transformation, 
it is defined on the coset space $G/H$. 
Further, since the action is first order in the $\tau$-derivative,
this coset space $G/H$ should be considered as the phase space.
For the case of $G = SU(3)$, the
phase space coordinates are then the Darboux variables $(\pi_i, \phi_i)$ in Eq.~(\ref{CGC6}).

Turning now to the quantum theory, we first note that the wave function
of a theory at time $\tau_{\rm f}$ can be defined by the
path integral
\begin{equation}
\psi (q_{\rm f}, \tau_{\rm f}) = \int [Dq] \,
e^{i S(q_{\rm f}, \tau_{\rm f}; q_{\rm i}, \tau_{\rm i}) }\,,
\label{A:syn2a}
\end{equation}
where we have used $q$ as a generic symbol for all the dynamical variables
and $S(q_{\rm f}, \tau_{\rm f}; q_{\rm i}, \tau_{\rm i})$ is the action 
over a path connecting $q_{\rm i}$ at $\tau_{\rm i}$ to
$q_{\rm f}$ at $\tau_{\rm f}$.
This shows that under a transformation on $q_{\rm f}$, the change
in the wave function is the same as the change in $e^{i S}$ 
evaluated at $\tau_{\rm f}$.
Thus from Eq.~(\ref{A:syn2}), we see that the wave functions for the
KKS action should have the property
\begin{equation}
\psi (g u ) = \psi (g) \, e^{-i w_s \theta_s }\,.
\label{A:syn3}
\end{equation}
This fixes the nature of the representation obtained by quantizing the action.

With this result, it is straightforward to write down the wave functions.
Consider the matrix element
\begin{equation}
\D^{(R)}_{ij} (g) = \bra{R, i} g \ket{R, j}
= \left[\exp\left( - i T_a \theta^a\right)\right]_{ij}, \hskip .3in
i,j = 1, 2, \cdots, {\rm dim}R\,,
\label{A:syn3a}
\end{equation}
which is the usual Wigner $\D$-function for the group.
Here $T_a$ are matrices in the Lie algebra in the representation $R$
of dimension ${\rm dim}R$.
We can also view this as the matrix element of the operator version
of $g = e^{- i {\hat T}_a \theta^a}$, with the states indicated in the previous  expression. We choose the state 
$\ket{R, w}$ which has the eigenvalues $h_s \ket{R, w} = w_s \ket{R, w}$.
Then $g u \ket{R, w} = g\ket{R, w} e^{- i w_s \theta_s}$ and we can identify
a basis for the wave functions as
\begin{equation}
\psi_i(g) = {C}\, \bra{R, i} g \ket{R, w} = {C}\, \D^{(R)}_{iw} (g)\,,
\label{A:syn3b}
\end{equation}
where ${C}$ is a normalization constant. From the orthogonality theorem on representations of a compact Lie group, 
we can see that ${C}$ can be taken as $\sqrt{{\rm dim} R}$.

There are a few more points of interest to us.
We can act on  the state $\ket{R, w}$ by raising and lowering type operators, which correspond to $E_{\talpha}$ and $E_{-{\talpha}}$ introduced in section \ref{sec:Goldstone}.
From the point of view of $G/H$ as a phase space, 
$E_{\talpha}$ and $E_{-{\talpha}}$ correspond to derivatives with respect to conjugate sets of Darboux coordinates. As is well-known, in the quantum theory, for reasons of irreducibility for the quantum algebra, the wave functions should only depend on half of the phase space coordinates.
In the present case, this can be enforced by requiring that 
$\psi$ should vanish upon the action of the raising operators $E_{\talpha}$ on $\ket{R, w}$.  (This is a consistent choice since the commutator $[E_{\talpha}, E_{\tilde\beta}] \sim E_{{\talpha} + {\tilde\beta}}$, without operators of the $E_{-{\talpha}}$ type.)
In other words, $\ket{R, w}$ should be the highest weight state for the representation $R$. With such a set of chosen $w$'s, this fixes the representation uniquely.

Secondly, the variables $\theta_s$ are angular variables, so a shift
of the form $\theta_s \rightarrow \theta_s + c_s (2\pi)$ will correspond to the same point on the group.\footnote{The constant $c_s$ will depend on the normalization of the generators. For example, for $SU(3)$, with $\Tr (t_a t_b ) = {1\over 2} \delta_{ab}$, we have $\theta_8 \rightarrow \theta_8 + \sqrt{12} \,2\pi$ for the periodicity condition.} With the property in Eq.~(\ref{A:syn3}), the single-valuedness
of $\psi$ will require certain quantization conditions on $w_s$.
These turn out to be exactly what is needed for a UIR of the group $G$.

There is another useful feature of this action.
Defining a set of charges
\begin{equation}
Q^a = 2\, \Tr \bigl( w_s h_s g^{-1} t^a g \bigr)\,,
\label{A:syn3c}
\end{equation}
it is easy to see that $Q^a$ are unchanged under $g \rightarrow g u$,
$u \in H$. Thus $Q^a$ are functions on the phase space $G/H$.
The canonical quantization of the KKS action then shows that
\begin{eqnarray}
Q^a \, \psi_i (g) &=& \psi_i(- t_a g) = - (T_a)_{ij} \psi_j (g)\,,\nonumber\\
{}[Q^a, Q^b] &=& i f_{abc} \, Q^c\,.
\label{A:syn3d}
\end{eqnarray}
The quantum version of $Q^a$ are thus matrices corresponding to the
generators of the Lie algebra.
We see that the KKS action represents a classical action which upon quantization leads to the required nonabelian algebra of charges.

We can use the KKS action and the representation of
$T_a$ arising from $Q^a$ as in Eq.~(\ref{A:syn3c}) upon quantization
to write down a path integral expression for the Wilson line.
For this, we consider the gauged version of the action in Eq.~(\ref{syn4a}),  
\begin{equation}
S(g, A) = 2  i \int d\tau~ \Tr  \bigl[ w_s h_s  g^{-1} \left( \partial_\mu g + (-it_a )A^a_\mu g\right)\bigr] \frac{dx^\mu}{d\tau}\,.
\label{A:syn4a}
\end{equation}
The $\tau$-derivative is along the curve $C$ given by
$x^\mu(\tau )$.
The first term, which is just the
KKS action in Eq.~(\ref{A:syn1}) is first order in the $\tau$-derivatives and 
therefore does not contribute to the Hamiltonian. 
The second term does lead to a nonzero 
Hamiltonian (for the degrees of freedom in $g$) given by\footnote{We consider here a fixed curve so that $x^\mu$ is not treated as a dynamical variable at this stage. Thus ${\mathcal H}$ is just the Hamiltonian for the
degrees of freedom in $g$.}
\begin{equation}
{\mathcal H} = - 2 \, \Tr\bigl[ w_s h_s  g^{-1} t_a g \bigr] A^a_\mu {dx^\mu \over d\tau}
= Q_a A^a_\mu {\dot x}^\mu\,,
\label{A:syn4b}
\end{equation}
where $Q_a$ is the classical color charge in Eq.~(\ref{A:syn3c}).
The quantum operator corresponding to ${\mathcal H}$ will have
$Q_a \rightarrow T_a$.
Therefore, the Wilson line operator can be expressed as
\begin{equation}
W_{ij}(R, C) = \left[ {\mathcal T} \exp\left( - i \int d\tau {\hat{\mathcal H}} \right)\right]_{ij}\,,
\label{A:syn4c}
\end{equation}
where ${\hat{\mathcal H}}$ is the operator version of ${\mathcal H}$.
Since $\tau$ varies long the path,
$\tau$-ordering (indicated by $\mathcal T$ above)  is the same as path-ordering.

We can now construct a path integral for this expression
by the usual process of splitting 
the range of $\tau$ into infinitesimal segments. 
The result is
\begin{equation}
W_{ij}(R, C) = \int [d\mu(g)] ~ \psi_i(g,1)^*  \,\left[ e^{i S (g, A)} \right]\, \psi_j(g,0)\,,
\label{A:syn5}
\end{equation}
where the action is that given in Eq.~(\ref{A:syn4a}). This is the central result that is used in the main text. 

In this action we see the classical version of the Hamiltonian as expected
for a path integral. The first term, namely the KKS action in Eq.~(\ref{A:syn1}), arises from expansion of terms like $\braket{g'}{g''}$ when using 
completeness relations. Indeed, the KKS action is the
equivalent of the familiar $p_i {\dot q}_i$-terms in Hamiltonian
path integrals.\footnote{As alluded to in the main text, an equivalent formulations exists in the worldine formalism, where the Wilson line can be represented by an worldline action~\cite{DHoker:1995aat} which, upon quantization, yields the nonabelian algebra of color charges~\cite{Mueller:2019gjj}. }
\section*{Appendix B: The lab frame and the
expansion of the Goldstone action}
\def\theequation{B\arabic{equation}}
\setcounter{equation}{0}
In the first part of this  appendix, we will consider the expression of the kinetic term for the Goldstone modes in the lab frame.
We will subsequently provide a detailed discussion of the  expansion of the kinetic term of the Goldstone fields
in Eq.~(\ref{syn18}).

If $u^\mu$ is the four-velocity of the condensate with respect to the lab frame, the relevant Lorentz transformation matrix 
$\Lambda^\mu_{~\nu}$ is given by
\begin{equation}
\Lambda^0_{~0} = u^0, \hskip .2in
\Lambda^0_{~i} = u^i, \hskip .2in \Lambda^i_{~0} = u^i,
\hskip .2in \Lambda^i_{~j} = \delta^i_j + {u^i u^j \over (1+ u^0)}\,.
\label{L1}
\end{equation}
The action in the rest frame of the condensate has the form
\begin{equation}
S_{\rm Goldstone} = -{1\over 2} f_G^2\int  \sigma^{\mu\nu} \sum_{\talpha} \Tr \left( E_{\bf\talpha}
U^{-1} \DU_\mu U\right) \Tr \left( E_{-\bf\talpha} U^{-1} \DU_\nu U\right)\,.
\label{L2}
\end{equation}
Therefore in the lab frame we have the action
\begin{equation}
S_{\rm Goldstone} = -{1\over 2} f_G^2\int  \sigma^{\lambda\sigma} 
\Lambda^\mu_{~\lambda} \Lambda^\nu_{~\sigma} \,\sum_{\talpha} \Tr \left( E_{\bf\talpha}
U^{-1} \DU_\mu U\right) \Tr \left( E_{-\bf\talpha} U^{-1} \DU_\nu U\right)\,.
\label{L3}
\end{equation}
Recall that $\sigma^{00} =1$, $\sigma^{ij} = - \delta^{ij} (1- \kappa)$. It is then easy to see that
\begin{equation}
   \sigma^{\lambda\sigma} 
\Lambda^\mu_{~\lambda} \Lambda^\nu_{~\sigma}
= \eta^{\mu\nu} - \kappa \left( \eta^{\mu\nu} - u^\mu u^\nu\right)\,.
\label{L4}
\end{equation}
In general, $u^\mu$ is an $X$-dependent flow velocity. 
If we have just one current for the Cartan subalgebra, we can take
$u^\mu = j^\mu/\sqrt{j^2}$.
For the color group $SU(3)$, we can have two currents
for the Cartan subalgebra. In defining a rest frame for the
condensate, we are assuming that there is no relative flow between
the $t_3$ and $t_8$ currents, so that their flow velocity in the lab frame is the same. In this approximation, we can take
$u^\mu = j_s^\mu/\sqrt{j_s^2}$ for any $s$.
The action in Eq.~(\ref{L3}) can thus be written using Eq.~(\ref{L4}) as
\begin{equation}
S_{\rm Goldstone} = -{1\over 2} f_G^2\int 
{1\over j^2_s} \left( \eta^{\mu\nu} (1- \kappa) j^2_s + \kappa j^\mu_s j^\nu_s
\right)
\sum_{\talpha} \Tr \left( E_{\bf\talpha}
U^{-1} \DU_\mu U\right) \Tr \left( E_{-\bf\talpha} U^{-1} \DU_\nu U\right)\,.
\label{L5}
\end{equation} 
The integration is over the region of support for the condensate.
If we allow for different flow velocities, this form of the expression should hold for each component of the Cartan subalgebra, so that the action may be written as
\begin{eqnarray}
S_{\rm Goldstone} &=& -{1\over 2} f_G^2\int 
\sum_s {1\over j^2_s} \left( \eta^{\mu\nu} (1- \kappa) j^2_s + \kappa j^\mu_s j^\nu_s
\right)
\sum_{\talpha} \Tr \left( E_{\bf\talpha}
U^{-1} \DU_\mu U\right) \Tr \left( E_{-\bf\talpha} U^{-1} \DU_\nu U\right)\nonumber\\
&=&-{1\over 2} f_G^2\int 
{\tilde \sigma}^{\mu\nu} 
\sum_{\talpha} \Tr \left( E_{\bf\talpha}
U^{-1} \DU_\mu U\right) \Tr \left( E_{-\bf\talpha} U^{-1} \DU_\nu U\right)\,.
\label{L6}
\end{eqnarray}
where
\begin{equation}
    {\tilde \sigma}^{\mu\nu} =
    \sum_s {1\over j^2_s} \left( \eta^{\mu\nu} (1- \kappa) j^2_s + \kappa j^\mu_s j^\nu_s
\right)\,.
\label{L7}
\end{equation}
\setlength{\belowdisplayskip}{3pt}

We will now expand this action in terms of the Goldstone fields and carry out 
this expansion to fourth order
in the fields. For $SU(3)$, we can identify
\begin{equation}
E_{\talpha} = \bigl( t_1 +i t_2, t_4 + i t_5, t_6 + i t_7\bigr),
\hskip .3in
E_{-{\talpha}} = \bigl( t_1 - i t_2, t_4 - i t_5, t_6 - i t_7\bigr)\,.
\label{B1}
\end{equation}
The action for the Goldstone modes is thus
\begin{equation}
S = - {1\over 2} f_G^2\int {\tilde \sigma}^{\mu\nu} \sum_A \Tr (E_A U^{-1} D_\mu U) \Tr(E_A U^{-1}D_\nu U)\,,
\label{B2}
\end{equation}
where $A = 1, 2, 4, 5, 6,7$. Further, from the Fierz identity, the generators obey the completeness relation
\begin{equation}
\sum_A (E_A)_{\alpha \beta} (E_A)_{\gamma\delta} =
{1\over 2} \delta_{\beta\gamma} \delta_{\alpha\delta} - {1\over 6} \delta_{\alpha\beta} \delta_{\gamma\delta}
- (t_3)_{\alpha\beta} (t_3)_{\gamma\delta} - (t_8)_{\alpha\beta}(t_8)_{\gamma\delta}\,.
\label{B3}
\end{equation}
We use this to simplify the action as
\begin{eqnarray}
S&=& - {1\over 2} f_G^2 \int {\tilde \sigma}^{\mu\nu} \left[ {1\over 2} \Tr (\DU_\mu U U^{-1} \DU_\mu U U^{-1})
- \Tr (h_s U^{-1} \DU_\nu U) \Tr (h_s U^{-1}\DU_\mu U) \right]\nonumber\\
&=& {1\over 8} f_G^2 \int {\tilde \sigma}^{\mu\nu}\left[ \delta_{ab} - U_{as} U_{bs} \right] I_\mu^a I_\nu^b\,.
\label{B4}\\
U_{as} &=& 2\, \Tr (h_s U^{-1} t_a U), \hskip .3in
\DU_\mu U U^{-1} = (-i t_a ) I_\mu^a\,,
\nonumber
\end{eqnarray}
where $s = 3,8$ and $a =1, 2, \cdots, 8$.
We now parametrize $U$ as $U = e^{ i \Phi}$, $\Phi = \phi^a t_a$.
Since we are interested in $SU(3)/[U(1) \times U(1)]$, we can restrict
the fields to the coset directions, namely, $\phi^3 = \phi^8 = 0$.
Notice that $U_{as}$ is the $(as)$-matrix element of the adjoint
representation of $U$, therefore
$U_{as} = \left( e^{i \phi^a  T_a}\right)_{as}$, where
$(T_a )_{bc} = -i f_{abc}$.
Thus we have
\begin{equation}
U_{as} = \delta_{as} + \phi^A f_{A a s} + {1\over 2}
\phi^A \phi^B f_{A a b} f_{B b s} + 
\cdots
\label{B5}
\end{equation}
We also have
\begin{eqnarray}
\DU U U^{-1} &=& dU U^{-1} + (-it_a)A^a\nonumber\\
&=& d(e^{i \Phi}) e^{-i \Phi}  + (-it_a)A^a = (-it_a) I^a_\mu\nonumber\\
I^a_\mu &=&  \left[ A^a_\mu - \partial_\mu \phi^a  + {1\over 2} f^{aBC} \phi^B \partial_\mu \phi^C + {1\over 6} f^{l BC} f^{l D a} \phi^B \partial_\mu\phi^C \phi^D + \cdots\right]\,.
\label{B6}
\end{eqnarray}
Recall that as discussed in Sec.~\ref{sec:Goldstone-Kinetic-term}, $\DU$ above is not the usual covariant derivative acting on vector fields but the left covariant derivative. 

We find
\begin{eqnarray}
I^a_\mu I^a_\nu &=& {1\over 2}\Bigl[A^s_\mu A^s_\nu + (A^A_\mu - \partial_\mu \phi^A)
 (A^A_\nu - \partial_\nu \phi^A)
+ {1\over 4} f^{s BC} f^{s DE} \phi^B \partial_\mu \phi^C \, \phi^D \partial_\nu \phi^E \nonumber\\
&&+ {1\over 4} f^{A BC} f^{A DE} \phi^B \partial_\mu \phi^C \, \phi^D \partial_\nu \phi^E + A^s_\mu f^{s BC} \phi^B \partial_\nu \phi^C
+ A^A_\mu f^{A BC} \phi^B \partial_\nu \phi^C \nonumber\\
&&+ {1\over 3} f^{A BC} f^{A Ds} \phi^B \partial_\mu \phi^C \phi^D A^s_\nu
+ {1\over 3} f^{l BC} f^{l DA} \phi^B \partial_\mu \phi^C \phi^D 
(A^A_\nu -\partial_\nu \phi^A) + ({\mu \leftrightarrow \nu})\Bigr]
\label{B7}
\end{eqnarray}
\begin{eqnarray}
U_{as} I^a_\mu &=& A^s_\mu + {1\over 2} f^{s BC} \phi^B \partial_\mu \phi^C
+ {1\over 6} f^{s AD} f^{ABC} \phi^B \partial_\mu\phi^C \phi^D
+ f^{s AB} \phi^A (A^B_\mu - \partial_\mu\phi^B)\nonumber\\
&&+ {1\over 2} f^{s DA} f^{ABC} \phi^B \partial_\mu \phi^C \phi^D
+ {1\over 2} \phi^A \phi^B f^{Aa C} f^{B C s} (A^a_\mu - \partial_\mu\phi^a)
+\cdots\nonumber\\
&=&A^s_\mu - {1\over 2} f^{s BC} \phi^B \partial_\mu \phi^C
+ f^{s BC} \phi^B A^C_\mu
+ {1\over 6} f^{s AD} f^{ABC} \phi^B \partial_\mu\phi^C \phi^D
+ {1\over 2} \phi^A \phi^B f^{Aa C} f^{B C s} A^a_\mu \,.
\label{B8}
\end{eqnarray}
Working out the product of two of these to quartic order in the fields we find
\begin{eqnarray}
(U_{as} I^a_\mu) (U_{as} I^a_\nu) &=&{1\over 2} \Bigl[ A^s_\mu A^s_\nu + {1\over 4} f^{s BC} f^{s DE}
 \phi^B \partial_\mu \phi^C \,\phi^D \partial_\nu \phi^E
 + f^{s BC}  f^{s DE} \phi^B A^C_\mu \phi^D A^E_\nu\nonumber\\
&& - f^{sBC} A^s_\mu \phi^B \partial_\nu \phi^C
 + 2 f^{sBC} A^s_\mu \phi^B A^C_\nu
 + {1\over 3} f^{s AD} f^{ABC}  A^s_\mu \phi^B \partial_\nu\phi^C \phi^D
 \nonumber\\
 && +f^{sBC} f^{AaC} A^s_\mu \phi^A \phi^B A^a_\nu 
 - f^{sBC} f^{sDE} \phi^B \partial_\mu\phi^C \phi^D A^E_\nu
 +(\mu \leftrightarrow \nu )
 + \cdots
 \label{B9}
\end{eqnarray}
From (\ref{B7}) and (\ref{B9}) we get
\begin{eqnarray}
{\tilde \sigma}^{\mu\nu}\left[ \delta_{ab} - U_{as} U_{bs} \right] I_\mu^a I_\nu^b
&=&{\tilde \sigma}^{\mu\nu} \Bigl[(\partial_\mu \phi^A - A^A_\mu)
(\partial_\nu \phi^A - A^A_\nu)
+ {1\over 4} f^{A BC} f^{A DE} \phi^B \partial_\mu \phi^C \, \phi^D \partial_\nu \phi^E + 2 f^{sBC} A^s_\mu \phi^B \partial_\nu \phi^C\nonumber\\
&&+ A^A_\mu f^{A BC} \phi^B \partial_\nu \phi^C
+ {1\over 3} f^{l BC} f^{l DA} \phi^B \partial_\mu \phi^C \phi^D 
(A^A_\nu -\partial_\nu \phi^A)\nonumber\\
&&- f^{s BC}  f^{s DE} \phi^B A^C_\mu \phi^D A^E_\nu
-2 f^{sBC} A^s_\mu \phi^B A^C_\nu
 -f^{sBC} f^{AaC} A^s_\mu \phi^A \phi^B A^a_\nu \nonumber\\
 &&+ f^{sBC} f^{sDE} \phi^B \partial_\mu\phi^C \phi^D A^E_\nu
 + \cdots\Bigr]
 \label{B10}
\end{eqnarray}
To get the standard normalization for the leading term, we define
$\phi^A = (2/f_G) {\varphi}^A$. For the action from Eq.~(\ref{B4}) we then find
\begin{eqnarray}
S&=& {1\over 2} \int {\tilde \sigma}^{\mu\nu}\biggl[
\left(\partial_\mu {\varphi}^A - {f_G \over 2} A^A_\mu\right)
\left(\partial_\nu {\varphi}^A - {f_G \over 2} A^A_\nu\right)
+ {1\over f_G^2} 
f^{A BC} f^{A DE} {\varphi}^B \partial_\mu {\varphi}^C \, {\varphi}^D \partial_\nu {\varphi}^E 
+2 f^{sBC} A^s_\mu {\varphi}^B \partial_\nu {\varphi}^C
\nonumber\\
&&~~+ A^A_\mu f^{A BC} \varphi^B \partial_\nu \varphi^C
- {4\over 3 f_G^2} f^{l BC} f^{l DA} \varphi^B \partial_\mu \varphi^C \varphi^D 
( \partial_\nu \varphi^A - {f_G \over 2} A^A_\nu)\nonumber\\
&&~~- f^{s BC}  f^{s DE} \varphi^B A^C_\mu \varphi^D A^E_\nu
- f_G f^{sBC} A^s_\mu \varphi^B A^C_\nu
 -f^{sBC} f^{AaC} A^s_\mu \varphi^A \varphi^B A^a_\nu \nonumber\\
 &&~~+ {2 \over f_G} f^{sBC} f^{sDE} \varphi^B \partial_\mu\varphi^C \varphi^D A^E_\nu
 + \cdots\biggr]\,.
 \label{B11}
\end{eqnarray}

\bibliography{reference_Glass-Fluid}




\end{document}